
\input phyzzx
\def\np{Nucl. Phys.}
\def\pl{Phys. Lett.}
\def\prl{Phys. Rev. Lett.}

\def\cmp{Comm. Math. Phys.}

\def\mpl{Mod. Phys. Lett.}
\def\half{{1\over 2}}

\def\ex{{\hbox{\rm e}}}

\def\ker{{\hbox{\rm Ker}}}

\def\nor{}
\def\norp{}

\tolerance=500000
\overfullrule=0pt

\def\mani{\cal{M}}

\def\ele{{\hbox{\sevenrm L}}}
\def\ere{{\hbox{\sevenrm R}}}
\def\zb{{\bar z}}

\tolerance=500000
\overfullrule=0pt
\pubnum={US-FT-9/91 \cr
     FTUAM-91/40 \cr}
\date={December 1991}
\pubtype={}
\titlepage

\title{TOPOLOGICAL MATTER IN TWO DIMENSIONS}
\author{J.M.F. Labastida\foot{E-mail: LABASTIDA@EUSCVX.DECNET}}
\address{Departamento de F\'\i sica de Part\'\i
culas\break Universidade de Santiago\break
E-15706 Santiago de Compostela, Spain}
\andauthor{P.M. Llatas}
\address{Departamento de F\'\i sica Te\'orica \break
Universidad Aut\'onoma de Madrid\break
E-28049 Madrid, Spain}

\abstract{Topological quantum field theories containing matter fields are
constructed by twisting $N=2$ supersymmetric quantum field theories. It is
shown that $N=2$ chiral (antichiral) multiplets lead to topological sigma
models while $N=2$ twisted chiral (twisted antichiral) multiplets lead to
Landau-Ginzburg type topological quantum field theories. In addition,
topological gravity in two dimensions is formulated using a gauge principle
applied to the topological algebra which results after the twisting of $N=2$
supersymmetry.}

\endpage

\chapter{\caps Introduction}

Recently, there has been a rapid development of topological gravity
\REF\lpw{J.M.F Labastida, M. Pernici and E. Witten\journal\np&B310(88)611}
\REF\ms{D. Montano and J. Sonnenschein\journal\np&B313(89)258}
\REF\mp{R.C. Myers and V. Periwal\journal\np&B333(90)536}
\REF\wiuno{E. Witten\journal\np&B340(90)281}
\REF\dist{J. Distler\journal\np&B342(90)523}
\REF\ver{E. Verlinde and H. Verlinde\journal\np&(91)457} [\lpw-\ver]
as well as conformal topological matter in two dimensions
\REF\ey{T. Eguchi and S.K. Yang\journal\mpl&A5(90)1693}
 [\ey]
coupled to topological gravity
\REF\li{K.Li, Topological Gravity and Minimal Matter, CALTECH Preprint,
CALT-68-1661; Recursion Relations in Topological Gravity with Minimal
Matter, CALTECH Preprint, CALT-68-1670}
\REF\dvv{R. Dijkgraaf, E. Verlinde
and H. Verlinde\journal\np&B352(91)59}
\REF\vw{E. Verlinde and N. Warner,
Topological Landau-Ginzburg Matter at $c=3$, IAS preprint, IASSNS-HEP-91/16}
[\li,\dvv,\vw].
 Since the relation between
topological gravity and one-matrix models
\REF\mm{E. Brezin and V. Kazakov\journal\pl&B236(90)144; M. Douglas and S.
Shenker\journal\np&B335(90)635; D.J. Gross and A.A.
Migdal\journal\prl&64(90)717} [\mm] was pointed out by Witten
\REF\widos{E. Witten, On the Structure of the Topological Phase of Two
Dimensional Gravity, IAS preprint, IASSNS-HEP-89/66} [\widos],
many efforts have been
carried out to analyze the scope of this type of relation
\REF\diwi{R. Dijkgraaf and E. Witten\journal\np&B342(90)486}
\REF\witres{E. Witten, On the Konsevich Model and Other Models of Two
Dimensional Gravity, IAS preprint, IASSNS-HEP-91/24, 1991} [\diwi,\witres].
Similarly, it has been shown that topological gravity coupled to some kind of
conformal topological matter is related to multimatrix models
\REF\verver{R. Dijkgraaf, E. Verlinde and H. Verlinde,
Topological Strings in $d<1$, Princeton preprint, PUPT-1204;
Loop Equations and Virasoro Constraints in Non-Perturbative 2D Gravity,
Princeton preprint, PUPT-1184; Notes on Topological String Theory and 2D
Quantum Gravity, Princeton preprint, PUPT-1217}
[\li,\verver]. Though the study of
topological matter in two dimensions is important in its own right, this
connection has stimulated the study of this type of topological quantum
field theories. Different aspects of conformal topological matter have been
considered in recent works
\REF\lerche{W. Lerche\journal\pl&B252(90)349}
\REF\rait{E.J. Raiten\journal\pl&B257(91)99}
\REF\blk{B. Blok and A. Varchenko, `Topological Conformal Field Theories and
the Flat Coordinates', IAS preprint, IASSNS-HEP-91/5, January 1991}
\REF\kaz{Y. Kazama, Novel Topological Conformal Algebras, Tokyo preprint,
UT-Komaba 90-30, January 1991} \REF\ehy{T. Eguchi, S. Hosono and S. Yang,
Hidden Fermionic Symmetry in Conformal Topological Field Theories', Santa
Barabara preprint, NSF-ITP-91-18i, January 1991}
 \REF\fusu{K. Fujikawa and H. Suzuki,
Topological Conformal Algebra and BRST Algebra in Non-Critical String
Theories, Kyoto Preprint, YITP/U-91-17, March 1991} \REF\noji{S.
Nojiri\journal\pl&B262(91)419} \REF\kuma{A. Kumar and J. Maharana,
Topological Field Theory from $N=3$ Superconformal Theory, Bhubaneswar
preprint, April 1991} \REF\gism{A. Giveon and D. Smit, Properties of
Superstring Vacua from (Topological) Landau-Ginzburg Models, Berkeley
preprint, LBL-30831, June 1991}
\REF\ceva{S. Cecotti and C. Vafa, Topological Anti-Topological Fusion,
Harvard preprint, HUTP-91/A031, June 1991} [\lerche-\ceva]. In this paper we
will study topological matter in two dimensions from a general point of view,
\ie, without restricting ourselves to the conformal case.

So far, all matter coupled to topological gravity has concerned
conformal topological matter resulting after twisting $N=2$ minimal models
\REF\wittres{E. Witten\journal\cmp&117(88)353} [\wittres,\ey]. These
theories have $c<3$ and therefore, after the twisting, they correspond to
topological gravity coupled to topological matter with $d<1$. In this sense,
the cases considered are below the $d=1$ barrier.  Presumably, beyond $d=1$
the topological and the ordinary phases of non-critical strings are very
different, contrary to what happens for $d<1$. However, it would be
interesting to understand the behavior of topological quantum field theories
for $d>1$ because, possibly, one could  gain some insight on the structure of
non-critical strings in more realistic dimensions. It is easy to find
topological matter which leads us beyond the $d=1$ barrier. For example,
some interesting models are the ones corresponding to topological
Wess-Zumino-Witten models.

In this paper we will set up the basis of the analysis of topological matter
from the point of view of twisting $N=2$ supersymmetric theories
\REF\luis{L. Alvarez-Gaum\'e and
D.Z. Freedman\journal\cmp&91(83)87}
\REF\roc{S.J. Gates, C.M. Hull and M.
Ro\v cek\journal\np&B248(84)157} [\luis,\roc]. We will
concentrate on  theories resulting from the twisting of
$N=2$ chiral (antichiral) multiplets and $N=2$ twisted chiral
(twisted antichiral) multiplets as the basic building multiplets of $N=2$
supersymmetry [\roc]. We will consider theories which contain only one type
of these multiplets. In future work we will enlarge our analysis to the case
in which one has  a mixture of these multiplets. This will lead to some
types of topological Wess-Zumino-Witten models. Of course, there are other
$N=2$ multiplets
\REF\tom{T. Buscher, U. Lindstrom and M.
Ro\v cek\journal\pl&202B(88)94} [\tom]
which should be also analyzed in this framework and that
presumably lead to a rich class of topological matter.

Three types of topological matter in two dimensions are known. The first
type, the topological sigma models, were found by Witten
\REF\wtsm{E. Witten\journal\cmp&118(88)411} [\wtsm]. He showed that they
correspond to some kind of twisted $N=2$ matter. In this work we will show
that this type of topological matter is the result of twisting $N=2$ chiral
multiplets. All topological matter constructed out of twisting $N=2$ chiral
multiplets will be referred as type A topological matter. The second type of
models, conformal topological field theories in two dimensions, were first
described from the point of view of their operator product algebra by Eguchi
and Yang [\ey]. Finally, the third type of  models,
also called topological Landau-Ginzburg models, were proposed by Vafa
\REF\vafa{C. Vafa\journal\mpl&A6(91)337} [\vafa].
In this paper we will show that this kind of topological matter is the
result of twisting twisted chiral multiplets. This topological matter will
be referred as type B topological matter. Our aim will be to construct the
most general class of models corresponding to each of the non-conformal types.
For type A it has been shown recently \REF\nos{J.M.F. Labastida and P.M.
Llatas,
Potentials for Topological Sigma Models, Santiago preprint, US-FT-8/91,
June 1991} [\nos] that the topological sigma models constructed by Witten
[\wtsm] can  be enlarged by adding potential terms for the case in
which the target manifold has isometries. As we will observe in
this paper, these potential terms are not of the usual type in $N=2$
supersymmetric theories (F-terms). Twisting $N=2$ supersymmetric theories
may lead to actions which are not gauge invariant. This is indeed what
occurs for the case of type A matter. F-terms are not allowed and the only
possibility to add potential terms to a topological sigma model is to make
use of the formulation of $N=2$ supersymmetric theories given by L.
Alvarez-Gaum\'e and D. Z. Freedman [\luis]. For type B topological matter,
however, F-terms are allowed and we will obtain the most general form of the
action corresponding to this type of topological matter. The simplest of our
models, the corresponding to a flat target space, will be identified with the
one constructed by Vafa [\vafa].

Though it is known that twisting $N=2$ supersymmetry one obtains
topological sigma models or topological Landau-Ginzburg models, it is not
clear which aspect (the type of twisting or the choice of $N=2$ supermultiplet)
is responsible for their difference. In this work we will show that
once the twisting procedure is fixed, different multiplets lead to
different topological models. Our analysis will provide the most general
action for either kind of topological matter. These results will set up the
framework for mixed multiplets which must be related to a topological
version of Wess-Zumino-Witten models.

Another important aspect of our work is the appearance of the
{\it topological algebra}. This algebra is a twisted version of $N=2$
supersymmetry which besides the Poincare algebra contains two basic odd
operators $Q$ and $G_{\mu}$. Indeed, the twisting of the spin 1/2
supersymmetric charges leads to two operators of spin 0 (one of them is $Q$)
and one operator of spin 1 (with two components). In the conformal case this
algebra  becomes the one first obtained in [\ey]. The topological algebra
provides a gauge principle to formulate topological gravity. Let us describe
how this works. Starting with an $N=2$ supersymmetric theory in a flat
two-dimensional space, one finds, after twisting, a theory whose action is
certainly invariant under transformations generated by the generators of the
algebra. However, if now one places the resulting theory on an arbitrary
two-dimensional manifold by introducing a metric $g_{\mu\nu}$, there is no
reason to expect the action to be invariant under the whole algebra. A
similar  situation with the $N=2$ supersymmetric case would lead to $N=2$
supergravity. It turns out that the resulting action is invariant under $Q$
for an arbitrary two-dimensional manifold while, in general, it is not
invariant under $G_{\mu}$. A simple way  to understand the
reason behind these facts is that $Q$ is a scalar operator and therefore one
does not expect complication when analyzed on a curved space. However,
$G_{\mu}$ is a vector operator and, as in ordinary gravity, general
covariance implies the introduction of additional fields. A twisted theory
by itself turns out to be a topological quantum field theory. On one hand
its action is $Q$-invariant. On the other hand, the topological algebra
almost guarantees that the energy-momentum tensor is $Q$-exact\foot{As it shown
in sect. 4, for type B topological matter this holds only on-shell.}. We will
introduce topological gravity by requiring the resulting action to be invariant
under the vector-like symmetry. This is obtained by gauging the symmetry
generated by $G_\mu$, \ie, by introducing an odd gauge field, $\psi_{\mu\nu}$,
which is a $Q$-partner of $g_{\mu\nu}$.

The way we obtain topological gravity by using a gauge principle applied to
the topological algebra is very reminiscent of the way gravity is induced
in string theory. In this way we could think of topological gravity
as induced topological gravity. Certainly, in this
construction all couplings to matter are automatically generated. What is
left is any additional term which involves just the gravity topological
multiplet. It seems that in two dimensions there is not a possible
non-trivial term of this kind to be added to the action. However, similar
procedures carried out in higher dimensions may lead to additional structures.
In principle, any term which is invariant under the transformations obtained
after the gauging is allowed. Notice that from this point of view one can think
of pure topological gravity as the case in which all matter is set to zero,
similarly to the case of string theory.

The paper is organized as follows. In sect. 2 we will discuss the general
form of the {\it topological algebra} which results after twisting $N=2$
supersymmetry. In sect. 3 we will carry out the twist corresponding to
chiral (antichiral) multiplets deriving the form of topological sigma
models. This type of topological matter will be referred as type A.  In
sect. 4 twisted chiral (twisted antichiral) multiplets are analyzed and
Landau-Ginzburg type of models are constructed.  The resulting type of
topological matter will be referred as type B. In sect. 5 we will construct
topological gravity using a gauge principle.
 Finally, in sect. 6 we state
our conclusions and we discuss the different  lines of investigation opened
in the framework of this paper. The appendix summarizes our conventions.

\endpage
\chapter{\caps The Topological Algebra}

In this section we will derive the general form of the algebra
corresponding to a topological quantum field theory (TQFT). We will derive
it by twisting $N=2$ supersymmetry but it applies to a more general set of
theories than the ones which can be obtained as the result of a twist.

The algebra of  $N=2$ supersymmetry in two space-time dimensions
is constituted by the following relations:
$$
\eqalign{
\{Q_{\alpha +}, {Q}_{\beta -}\} =&
\gamma^\mu_{\alpha\beta} P_\mu, \cr
\{Q_{\alpha +}, {Q}_{\beta +}\} =&\{Q_{\alpha -}, {Q}_{\beta -}\} =0,\cr
[Q_{\alpha a},P_\mu]=&[P_\mu,P_\nu]= 0,\cr
[J,Q_{\pm a}]=&\pm{1\over 2}Q_{\pm a},\cr}
\qquad\qquad
\eqalign{
[R,Q_{\alpha \pm}]=&\pm{1\over 2}Q_{\alpha \pm},\cr
[J,P_\mu]=&-i\epsilon_{\mu}{}^{\nu}P_\nu,\cr
[R,P_\mu]=&0,\cr
[J,R]=&[J,J]=[R,R]=0,\cr}
\eqn\cuno
$$
where $J$ is the generator of Lorentz $SO(2)$ transformations and
$R$ the generator of the internal $SO(2)$ symmetry.  In \cuno\ greek indices
from the beginning of the alphabet denote Lorentz $SO(2)$ spinor
representations, and latin indices spinor representations of the internal
$SO(2)$ symmetry. Greek indices from
the middle of the alphabet denote Lorentz vector representations.
The epsilon
symbol in this expression is taken in such a way that
$\epsilon^{12}=-\epsilon^{21}=1$, and $\gamma_\mu$ are  Dirac
gamma matrices. Throughout this paper we use two-dimensional Euclidean
space. Our conventions are summarized in the appendix. There are $N=2$
models where  $R$-symmetry is broken by potential terms. Of course, all
$N=2$ superconformal models are $R$-invariant. On the other hand, as it is
described below, there are $N=2$ models which are not superconformal
invariant and are $R$-invariant.
 From \cuno\  is clear that if one is able to twist in such a way that some
of the supersymmetric charges  becomes scalar and the other ones components
of a vector one possesses a momentum operator which is $Q$-exact.
Furthermore, it seems plausible that in addition the new scalar
charge squares to zero.

To carry out the twist we will perform  a change in the spin of the
supersymmetric charges. To do this we have to redefine the Lorentz
generator in such a way that under the new one some of the
supersymmetric charges behave as scalars. There are two obvious but
equivalent
 possibilities to carry this out. Certainly, by adding or subtracting
the generators $R$ and $J$ one finds that some of the operators
$Q_{\alpha a}$ become scalars respect to the resulting angular momentum
generator. Let us define, for example,
$$
\tilde J = J+R.
\eqn\cdos
$$
Respect to the new Lorentz generator $\tilde J$ one finds that $Q_{+,-}$
and $Q_{-,+}$ behave as scalars while the pair $Q_{+,+}$, $Q_{-,-}$ behave
as a vector. Let us make the following definitions to make manifest the
new Lorentz structure of each of the generators:
$$
\eqalign{
Q_\ele &= Q_{+,-}, \cr
Q_\ere &= Q_{-,+}, \cr
Q_{+,+} &= \gamma_{++}^\mu G_\mu, \cr
Q_{-,-} &= \gamma_{--}^\mu G_\mu. \cr}
\eqn\ccuatro
$$
Notice
that Lorentz $SO(2)$ and  internal $SO(2)$ indices are separated by a
comma when specified explicitly. Clearly, from \cuno\ follows that
$$
Q_\ele^2 =Q_\ere^2=\{Q_\ele,Q_\ere\}=0.
\eqn\ccinco
$$
Let us rewrite the algebra \cuno\ in terms of the new Lorentz generator
$\tilde J$ and the following operators defined from \ccuatro,
$$
Q=Q_\ele+Q_\ere,\,\,\,\,\,\,\,\,\,\,\,\,\,\,\,
M=Q_\ele-Q_\ere.
\eqn\cseis
$$
It turns out that the resulting algebra takes the following form,
$$
\eqalign{
Q^2 & = M^2 = \{Q,M\}=[Q,P_\mu] =[M,P_\mu] = 0, \cr
\{Q,G_\mu\} &= P_\mu, \cr
[Q,\tilde J] &=[M,\tilde J]=0, \cr
\{M, G_\mu\}& = -i\epsilon_\mu{}^\nu P_\nu, \cr
[\tilde J, P_\mu] &= -i\epsilon_\mu{}^\nu P_\nu, \cr
[\tilde J, G_\mu] &= -i\epsilon_\mu{}^\nu G_\nu, \cr
[P_\mu,P_\nu]&=\{G_\mu,G_\nu\}=[\tilde J,\tilde J] =0.\cr}
\eqn\csiete
$$
This algebra contains the ingredients of a TQFT.
It certainly possesses the generators of the ordinary Poincare group plus
a nilpotent operator, $Q$. Furthermore, the momentum operator is $Q$-exact.
In addition, the generators $G_\mu$ and $M$ can be thought as a kind of odd
version of the Poincare group. Notice however that the operator $M$ which
could play the role of odd Lorentz generator rotates $G_\mu$ into $P_\mu$.
This is on the other hand rather natural since it would be inconsistent with
the odd nature of $G_\mu$ and $M$ to rotate it into $G_\mu$.

It is worth to mention that in addition one has the $R$ generator. Its
action on the operators entering in the algebra \csiete\ is,
$$
\eqalign{
[R,Q] & = -{1\over 2}M,\cr
[R,M] & =  -{1\over 2}Q, \cr
[R,G_\mu] & =  -{i\over 2}\epsilon_\mu{}^\nu G_\nu, \cr
[R,R]&=[R,\tilde J] = [R,P_\mu] = 0.\cr}
\eqn\cocho
$$
 The algebra \csiete\ together with the relations \cocho\ constitute what
we will call {\it topological algebra}. In the coming section we will find
out that the $R$ symmetry can be redefined so that it can be regarded as
``ghost number". Actually, it is possible to introduce  this ghost number
symmetry in a more general framework using the fermion-number symmetry of the
$N=2$ supersymmetric theory. If $F$ is the generator of this symmetry, its
action on the operators entering the $N=2$ supersymmetric algebra \cuno\ is,
$$
\eqalign{
\{F,Q_{+,-}\}=& Q_{+,-},\cr
\{F,Q_{-,+}\}=& Q_{-,+},\cr}
\qquad\qquad
\eqalign{
\{F,Q_{+,+}\}=& -Q_{+,+},\cr
\{F,Q_{-,-}\}=& -Q_{-,-},\cr}
\eqn\extramas
$$
while it commutes with all other operators in \cuno. Notice that the
first relation in \cuno\ is consistent with \extramas\ since
$\gamma_{+-}^\mu P_\mu = 0$ (see the appendix).
 After the twisting one finds that,
$$
\{F,Q\} = Q, \,\,\,\,\,\,\,\,\,
\{F,M\} = M, \,\,\,\,\,\,\,\,\,
\{F,G_\mu\} = -G_\mu,
\eqn\extramasd
$$
while its action on all other generators in \csiete\ is trivial.
Clearly, the natural interpretation of these transformations in the
twisted theory is the corresponding to ``ghost number". The reason
being that one would think that the generator
$G_\mu$  lowers the ghost number by one unit while the
$Q$-generator increases it by the  same amount. This is consistent with the
second relation in \csiete. Notice that $F$ treats both components of
$G_\mu$ in the same footing while, according to \cocho, $R$ treats
differently right and left components.

Before constructing the models of topological matter let us review the
general features of a topological quantum field theory from the perspective
of the topological algebra \csiete.
First, we will recall what is understood by a topological quantum field
theory. Let us consider a quantum field theory defined on a manifold
$\mani$ endowed with a metric $g_{\mu\nu}$. This quantum field
theory is topological if there exist some correlation functions involving some
of the fields of the theory,
$\langle\phi_{i_1}\phi_{i_2}...\phi_{i_n}\rangle$ such that:
$$
{\delta\over \delta g^{\mu\nu}}
\langle\phi_{i_1}\phi_{i_2}...\phi_{i_n}\rangle=0.
\eqn\uno
$$
Here the indices of the field denote certain quantum numbers as well as,
possibly, space-time points, curves or surfaces.
If one considers only fields $\phi_i$ which are independent of the metric
$g_{\mu\nu}$,
one way to ensure a property like \uno\ in a quantum field theory is the
following. Let us assume that the theory possesses a symmetry and that the
fields $\phi_i$ entering into the correlation functions above are invariant
under such a symmetry,
$$
\delta\phi_i=0.
\eqn\dos
$$
Property \uno\ follows if in addition there exist a tensor $G_{\mu\nu}$
such that the energy-momentum tensor of the theory, $T_{\mu\nu}$, can
be written as
$$
T_{\mu\nu}=\delta G_{\mu\nu}.
\eqn\tres
$$
To verify this notice that,
$$
\eqalign{
{\delta\over \delta g^{\mu\nu}}&
\langle\phi_{i_1}\phi_{i_2}...\phi_{i_n}\rangle=
\langle\phi_{i_1}\phi_{i_2}...\phi_{i_n}T_{\mu\nu}\rangle\cr &=
\langle\phi_{i_1}\phi_{i_2}...\phi_{i_n}\delta G_{\mu\nu}\rangle=
\langle\delta(\phi_{i_1}\phi_{i_2}...\phi_{i_n}G_{\mu\nu})\rangle=0,
\cr}
\eqn\cuatro
$$
where in the last step we have used the fact the theory is invariant under the
symmetry. For example, if one possesses a lagrangian formulation, this last
statement means that the action and the functional integral measure are
invariant under such a symmetry. Notice that in this analysis there is no
need for $\delta$ to be a nilpotent transformation as is typically the case
in topological quantum field theories. In fact, we will find in sect. 3
some kind of topological matter which possesses an operator $Q$ which is
not nilpotent. Its origin from the point of view of twisting $N=2$
supersymmetry corresponds to the fact that the $N=2$ supersymmetric model
possesses central charges. In general, twisting $N=2$ theories with central
charges will provide realizations of topological quantum field theories
with $Q^2\neq 0$.

We will consider the symmetry \dos\ as the one generated by $Q$ in the
topological algebra \csiete. Notice that this is consistent with \tres.
Although \tres\ is stronger than $\{Q,G_\mu\}= P_\mu$, we will find that in
the models obtained after twisting $N=2$ supersymmetry \tres\ holds at least
on-shell. Condition \dos\ implies that the physical states of the theory, \ie,
the ones associated to topological invariants must satisfy,
$$
Q|\Psi\rangle=0.
\eqn\catorce
$$
Two states which differ by a $Q$-exact state must be identified since from
\csiete\ one has $Q^2=0$. In other words, physical states correspond to
cohomology classes of $Q$.
Once we have a state satisfying \catorce\ we may use the operator $G_\mu$
to create its partners.
The simplest partner consists of
$$ \int_{\gamma_1} G_{\mu} |\Psi\rangle
\eqn\quince $$
where $\gamma_1$ is a 1-cycle. One can easily verify using \csiete\ that
this new state satisfies \catorce:
$$
Q \int_{\gamma_1} G_{\mu} |\Psi\rangle =
\int_{\gamma_1} \{Q,G_{\mu} \}|\Psi\rangle =
\int_{\gamma_1} P_{\mu} |\Psi\rangle =0.
\eqn\dseis
$$
Similarly, one may construct other invariants tensoring $n$ operators
$G_\mu$ and integrating over $n$-cycles $\gamma_n$:
$$
\int_{\gamma_n} G_{\mu_1}G_{\mu_2}...G_{\mu_n} |\Psi\rangle.
\eqn\dsiete
$$
Notice that since the operator $G_\mu$ is odd the integrand in this
expression is an $n$-form. It is straightforward to prove that these
states also satisfy  condition \catorce. Therefore, starting with a
state  $|\Psi\rangle\in \ker Q$ we have built a set of partners or
descendants constructing a topological multiplet. The members of a
multiplet have well defined ``ghost" number. If one assigns ghost
number $-1$ to the operator $ G_{\mu}$ the state in \dsiete\ has ghost
number $-n$ plus the ghost number of $|\Psi\rangle$. Of course, $n$
is bounded by the dimension of the manifold $\mani$. Among the
states constructed in this way there may be many which are related via
another state which is $Q$-exact, \ie, which can be written as $Q$ acting
on some other state. Let us try to single out representatives at each
level of ghost number in a given topological multiplet.

Consider an $(n-1)$-cycle which is the boundary of an $n$-dimensional
surface, $\gamma_{n-1}=\partial S_n$. If one tried to build a state taking
such a cycle one would have ($P_\mu=\partial_\mu$),
$$
\int_{\gamma_{n-1}} G_{\mu_1}G_{\mu_2}...G_{\mu_{n-1}}
|\Psi\rangle= \int_{S_n} P_{[\mu_1}
G_{\mu_2}G_{\mu_3}...G_{\mu_{n}]}|\Psi\rangle=
Q \int_{S_n} G_{\mu_1}
G_{\mu_2}...G_{\mu_{n}}|\Psi\rangle,
\eqn\docho
$$
\ie, it is $Q$-exact.
The symbols [ ] in \docho\ indicate that all indices between them must by
antisymmetrized. In \docho\ use has been made of  \csiete.
This result tells us that  the representatives we are looking for are
built out of the homology cycles of the manifold $\mani$. Given a
manifold $\mani$, the homology cycles are equivalence classes among cycles,
the equivalence relation being that two $n$-cycles are equivalent if they
differ by a cycle which is the boundary of an $n+1$ surface. Thus,
knowledge on the homology of the manifold on which the TQFT is defined
allows us to classify the representatives among the operators \dsiete.
Let us assume that $\mani$ has dimension $d$ and that its homology cycles
are $\gamma_{i_n}$, $i_n=1,...,d_n$, $n=1,...,d$, being $d_n$ the dimension
of the  $n$-homology group. Then, the non-trivial partners or
descendants of a given $|\Psi\rangle$ ``highest-ghost-number" state are
labeled in the following way:
 $$
\int_{\gamma_{i_n}} G_{\mu_1}G_{\mu_2}...G_{\mu_n} |\Psi\rangle,
\,\,\,\,\,\,\,\,\,i_n=1,...,d_n,\,\,\,\,\,\,\, n=1,...,d.
\eqn\dnueve
$$

A similar construction to the one just described can be made for fields.
Starting with a field $\phi(x)$ which
satisfies,
$$
[Q,\phi(x)]=0
\eqn\veinte
$$
one can construct other fields using the operators $G_{\mu}$. These
fields, which we will call partners or descendants are antisymmetric
tensors defined as,
$$
\phi^{(n)}_{\mu_1\mu_2...\mu_n}(x)={1\over
n!}[G_{\mu_1},[G_{\mu_2}...[G_{\mu_n},\phi(x)\}...\}\},
\,\,\,\,\,\,\,\, n=1,...,d.
\eqn\veintep
$$
Using \csiete\ and \veinte\ one finds that these fields satisfy the
so-called ``topological descent equations":
$$
d \phi^{(n)} =  [Q,\phi^{(n+1)}\}
\eqn\vseis
$$
where the subindices of the forms
has been suppressed for simplicity, and the highest-ghost-number field
$\phi(x)$
    has been denoted
as $\phi^{(0)}(x)$. These equations enclose all the
relevant properties of the observables which are constructed out of them.
As we will see in subsequent sections these equations are very useful
to build the observables of the theory.
Let us consider an $n$-cycle and the following quantity:
$$
W^{(\gamma_n)}_\phi = \int_{\gamma_n} \phi^{(n)}.
\eqn\vsiete
$$
The subindex of this quantity denotes the highest-ghost-number state
out of which the form $\phi^{(n)}$ is generated. The superindex denotes
the order of such a form as well as the cycle which is utilized in the
integration. Using the topological descent equations \vseis\ it is
immediate to prove that $W^{(\gamma_n)}_\phi$ is indeed an observable
$$
[Q, W^{(\gamma_n)}_\phi\} =\int_{\gamma_n} [Q,\phi^{(n)}\}
=\int_{\gamma_n} d\phi^{(n-1)}=0.
\eqn\vocho
$$
Furthermore, if $\gamma_n$ is a trivial homology cycle, $\gamma_n=\partial
S_{n+1}$, one obtains that $W^{(\gamma_n)}_\phi$ is $Q$-exact,
$$
W^{(\gamma_n)}_\phi = \int_{\gamma_n} \phi^{(n)}=
\int_{S_{n+1}} d \phi^{(n)} =  \int_{S_{n+1}} [Q, \phi^{(n+1)}\}
=[Q,\int_{S_{n+1}}\phi^{(n+1)}\},
\eqn\vnueve
$$
and therefore its vacuum expectation value vanishes. Thus, similarly to
the previous analysis leading to \dnueve\ the observables of the theory
are operators of the form \vsiete:
$$
W^{(\gamma_{i_n})}_\phi,\,\,\,\,\,\, i_n=1,...,d_n,\,\,\,\,\,\, n=1,...,d,
\eqn\treinta
$$
where, as before, $d_n$ denote the dimension of the $n$-homology group.
Of course, these observables are a basis of observables but one can consider
arbitrary products of them leading to new ones.

The main goal of this paper is to apply the twisting procedure just
described  to some $N=2$ supersymmetric matter. We will analyze the two
basic $N=2$ supersymmetric multiplets, the chiral (antichiral) multiplet, and
the twisted (not to be confused with the twisting just explained regarding
the construction of TQFT out of $N=2$ supersymmetric theories) chiral
(twisted antichiral) multiplet. These multiplets are better described if one
starts with its definition in $N=2$ superspace. Let us therefore consider
$N=2$ superspace on which one has superspace covariant derivatives
$D_{\alpha a}$ satisfying the following algebra:
$$
\eqalign{
\{D_{+,+}\,\, ,D_{+,-}\} &= 2\partial_z, \cr
\{D_{-,-}\,\, ,D_{-,+}\} &= 2\partial_\zb, \cr}
\eqn\cincuenta
$$
while all other anticommutators among the $D_{\alpha a}$ vanish.

The two basic $N=2$ multiplets are  described by a scalar
$N=2$ superfield $\Phi$ satisfying the following relations:
$$
\eqalign{
D_{+,-}\Phi &= D_{-,-}\Phi =0, \,\,\,\,\,\,\,\,\,{\hbox{\tenrm
chiral}},\cr D_{+,-}\Phi &= D_{-,+}\Phi =0,
\,\,\,\,\,\,\,\,\,{\hbox{\tenrm twisted chiral}}.\cr}
\eqn\ciuno
$$
Of course, there exist also the antichiral and the twisted antichiral
versions of these multiplets,
$$
\eqalign{
D_{+,+}\hat\Phi &= D_{-,+}\hat\Phi = 0, \,\,\,\,\,\,\,\,\,{\hbox{\tenrm
antichiral}},\cr D_{+,+}\hat\Phi & = D_{-,-}\hat\Phi =0,
\,\,\,\,\,\,\,\,\,{\hbox{\tenrm twisted antichiral}}.\cr}
\eqn\cidos
$$
The two kinds of multiplets that we are going to treat lead, after
twisting, to two different types of TQFT. They have different content and
they allow different potential terms. We will refer to them as type A and
type B topological matter. They will be described in the following
sections.

\endpage

\chapter{\caps Type A Topological Matter}

In this section we will carry out the construction of theories involving
topological matter of type A. We will describe this construction in full
detail for this type of topological matter. The procedure concerning other
types of topological matter is similar. In the next section we
will concentrate mainly on the results concerning type B topological matter.

Let us consider a
collection of chiral superfields $X^I$ and the corresponding set of
antichiral superfields $X^{\bar I}$, ($I,\bar I=1,...,d$),
$$
D_{\alpha,-} X^I = 0,\,\,\,\,\,\,\,\,\,\, D_{\alpha,+}
X^{\bar I} =0.
\eqn\citres
$$
We will consider actions in $N=2$ superspace that involve these
sets of fields containing a non-chiral kinetic term (sometimes called
D-term) and a chiral superpotential term (sometimes called F-term).
The action takes the form,
 $$
S={\nor} \int d^2z d^4\theta K(X^I,X^{\bar I})
+\int d^2z \big( d^2\theta W(X^I) + \widehat{d^2\theta} \bar
W(X^{\bar I})\big), \eqn\cicuatro
$$
The quantities $X^I$, $X^{\bar I}$  can be thought of as
coordinates of a $2d$-dimensional Kahler manifold $M$ where  $K$ is the
Kahler potential. In this sense, the superpotential $W(X^I)$ ($\bar
W(X^{\bar I})$) can be thought of as a holomorphic (antiholomorphic)
scalar function on such a manifold.

The odd parts of the measure in both terms of the action are such that
when projecting into components,
$$
\eqalign{
d^4\theta &\rightarrow  D_{-,+}D_{+,+}D_{-,-}D_{+,-},\cr
d^2\theta &\rightarrow  D_{+,+}D_{-,+},\cr
\widehat{d^2\theta} &\rightarrow  D_{-,-}D_{+,-}.\cr }
\eqn\cicinco
$$
This indicates that when twisting to obtain the corresponding TQFT we can
not allow the superpotential term since the measure is not Lorentz
invariant respect to the new Lorentz generator $\tilde J$. Recall from
\cdos\ that $\tilde J$ is such that while $D_{-,+}$ behaves as a scalar,
$D_{+,+}$ behaves as the component of a vector. Therefore, we may have
only the D-term for the case of type A topological matter. As we will
discuss in sect. 4, for the case of type B topological
matter this kind of term is allowed and one is able to build a TQFT
containing potential terms. This is not however the end of the story
regarding potential terms for type A topological matter. It is well
known that there are some $N=2$ supersymmetric matter models which
contain potential terms and are such that they can not be derived
from an $N=2$ superspace formulation [\luis]. One may wonder if starting
from those models one can construct additional TQFT. In particular,
if those models are able to provide potential terms for type A
topological matter. In [\nos] we have  answered this question.
It turns out that indeed, those models provide potential terms for type A
topological matter but do not add any additional structure for the case of
type B. Notice also that the F-term in \cicuatro\ breaks $R$-symmetry.
However, since we are not allowed to have that term in the topological model,
$R$-invariance will be present. As we will discuss below, both type A and
type B topological matter possess $R$-symmetry. $R$-invariance must be a
feature of all TQFT which are constructed by twisting a Lorentz
invariant supersymmetric theory since $R=\tilde J - J$ and in that
case both Lorentz symmetries are preserved. At the end of this section we
will make a brief summary of the results obtained in [\nos] for
potential terms in type A topological matter. For the moment, however, we
will restrict ourselves to consider the model \cicuatro\ without
F-term.

Let us define component fields in the following way,
$$
\eqalign{
X^I| &=x^I, \cr
D_{+,+} X^I| & = \psi_{+,+}^I, \cr
D_{-,+} X^I |& = \psi_{-,+}^I, \cr
D_{-,+} D_{+,+} X^I| &= F_{-+,++}^I, \cr}
\qquad\qquad
\eqalign{
X^{\bar I}| &=x^{\bar I}, \cr
D_{+,-} X^{\bar I}| & = \psi_{+,-}^{\bar I}, \cr
D_{-,-} X^{\bar I}| & = \psi_{-,-}^{\bar I}, \cr
D_{+,-}D_{-,-} X^{\bar I}| &= F_{+-,--}^{\bar I}. \cr}
\eqn\cisiete
$$
Before proceeding with the twisting to obtain the corresponding TQFT let
us work out the supersymmetry transformations of the component fields.
This will be very useful because it will allow us to determine the
transformations under the symmetries of the topological algebra
\csiete\ and \cocho. The supersymmetry transformations are easily
obtained taking into account that for an $N=2$ superfield such a
transformation takes the form,
$$
\delta \Phi = \eta^{\alpha a} Q_{\alpha a} \Phi,
\eqn\cinueve
$$
where $\eta^{\alpha a}$ is a constant $N=2$ supersymmetry parameter.
Taking components in this transformation law and  using the definitions
\cisiete\ one finds,
$$
\eqalign{
\delta x^I & = \eta^{+,+}\psi_{+,+}^I+\eta^{-,+}\psi_{-,+}^I, \cr
\delta \psi_{+,+}^I &=\eta^{-,+} F_{-+,++}^I+2\eta^{+,-}\partial_z
x^I,\cr
\delta \psi_{-,+}^I &=-\eta^{+,+} F_{-+,++}^I+2\eta^{-,-}\partial_\zb
x^I,\cr
\delta  F_{-+,++}^I&=2\eta^{-,-} \partial_\zb\psi_{+,+}^I -2
                        \eta^{+,-} \partial_z \psi_{-,+}^I,\cr
\delta x^{\bar I} & = \eta^{-,-}\psi_{-,-}^{\bar
I}+\eta^{+,-}\psi_{+,-}^{\bar I}, \cr
\delta \psi_{-,-}^{\bar I}
&=\eta^{+,-} F_{+-,--}^{\bar I}+2\eta^{-,+}\partial_\zb x^{\bar I},\cr
\delta \psi_{+,-}^{\bar I} &=-\eta^{-,-} F_{-+,--}^{\bar
I}+2\eta^{+,+}\partial_z x^{\bar I},\cr
\delta  F_{+-,--}^{\bar I}&=2\eta^{+,+} \partial_z\psi_{-,-}^{\bar
I} -2\eta^{-,+} \partial_\zb \psi_{+,-}^{\bar
I}.\cr} \eqn\sesenta
$$
The transformations under the $R$-symmetry in \cuno\ are obvious from the
$SO(2)$ indices carried out by the fields.

After this detailed description of the $N=2$ supersymmetric model we are
ready to carry out the twist. Under the new Lorentz generator $\tilde J$
in \cdos\ the Lorentz structure of each of the fields is simple to find
out since one just have to think of the indices as all belonging to the
same $SO(2)$. To make manifest the new Lorentz structure we will make
the following definitions:
$$
\eqalign{
\chi^I &= \psi_{-,+}^I,\cr
\chi^{\bar I} &= \psi_{+,-}^{\bar I},\cr}
\qquad
\eqalign{
\rho_z^I &= \psi_{+,+}^I,\cr
\rho_\zb^{\bar I} &= \psi_{-,-}^{\bar I},\cr}
\qquad
\eqalign{
F_z^I &= F_{-+,++}^I,\cr
F_\zb^{\bar I} &=  F_{+-,--}^{\bar I}.\cr}
\eqn\suno
$$
 After these definitions the $R$-transformations
of the fields are not manifest anymore. Let us summarize them here: $$
\eqalign{
[R,x^I] & = 0, \cr
[R,\chi^I] & = {1\over 2} \chi^I, \cr
[R,\rho^I_z] & = {1\over 2} \rho^I_z, \cr
[R,F^I_z] & =  F^I_z, \cr}
\qquad\qquad
\eqalign{
[R,x^{\bar I}] & = 0, \cr
[R,\chi^{\bar I}] & = -{1\over 2} \chi^{\bar I}, \cr
[R,\rho^{\bar I}_\zb] & = -{1\over 2} \rho^{\bar I}_\zb, \cr
[R,F^{\bar I}_\zb] & =  -F^{\bar I}_\zb. \cr}
\eqn\extrae
$$
Ghost numbers for the fields \suno\ are easily obtained from the
projections \cisiete\ and the fact that the ghost numbers of the
superspace covariant derivatives are the same as the ones of the
corresponding supersymmetry generators \extramas. Assuming that the
superfields $X^I$, $X^{\bar I}$ have ghost number 0, one finds,
$$
\eqalign{
[F,x^I] & = 0, \cr
[F,\chi^I] & =  \chi^I, \cr
[F,\rho^I_z] & = - \rho^I_z, \cr
[F,F^I_z] & =  0, \cr}
\qquad\qquad
\eqalign{
[F,x^{\bar I}] & = 0, \cr
[F,\chi^{\bar I}] & = \chi^{\bar I}, \cr
[F,\rho^{\bar I}_\zb] & = -\rho^{\bar I}_\zb, \cr
[F,F^{\bar I}_\zb] & =  0. \cr}
\eqn\extramast
$$

The transformations of
the fields under the generators $Q$, $M$ and $G_\mu$ of the
topological algebra \csiete\ follow from \sesenta:
$$ \eqalign{
[Q,x^I]& = \chi^I, \cr
[M,x^I] & = -\chi^I, \cr
[G_z,x^I] & = {1\over 2}\rho_z^I, \cr
[G_\zb,x^I] & = 0,\cr}
\qquad\qquad
\eqalign{
[Q,x^{\bar I}] & = \chi^{\bar I}, \cr
[M,x^{\bar I}] & = \chi^{\bar I}, \cr
[G_z,x^{\bar I}] & = 0, \cr
[G_\zb,x^{\bar I}] & = {1\over 2}\rho_\zb^{\bar I},\cr}
\eqn\sdos
$$
$$
\eqalign{
\{Q,\chi^I\} & = 0,\cr
\{M,\chi^I\} & = 0,\cr
\{G_z,\chi^I\} & = -{1\over 2}F_z^I,\cr
\{G_\zb,\chi^I\} & = \partial_\zb x^I,\cr}
\qquad\qquad
\eqalign{
\{Q,\chi^{\bar I}\} & = 0,\cr
\{M,\chi^{\bar I}\} & = 0,\cr
\{G_z,\chi^{\bar I}\} & = \partial_z x^{\bar I},\cr
\{G_\zb,\chi^{\bar I}\} & = -{1\over 2}F_\zb^{\bar I},\cr}
\eqn\stres
$$
$$
\eqalign{
\{Q,\rho_z^I\} & = 2\partial_z x^I + F_z^I,\cr
\{M,\rho_z^I\} & = 2\partial_z x^I - F_z^I,\cr
\{G_z,\rho_z^I\} & = 0,\cr
\{G_\zb,\rho_z^I\} & =0,\cr}
\qquad\qquad
\eqalign{
\{Q,\rho_\zb^{\bar I}\} & = 2\partial_\zb x^{\bar I} +
    F_\zb^{\bar I},\cr
\{M,\rho_\zb^{\bar I}\} & = -2\partial_\zb x^{\bar I} +
    F_\zb^{\bar I},\cr
\{G_z,\rho_\zb^{\bar I}\} & = 0,\cr
\{G_\zb,\rho_\zb^{\bar I}\} & =0,\cr}
\eqn\scuatro
$$
$$
\eqalign{
[Q,F_z^I] & = -2\partial_z\chi^I, \cr
[M,F_z^I] & = -2\partial_z\chi^I, \cr
[G_z,F_z^I] & = 0, \cr
[G_\zb,F_z^I] & = \partial_\zb\rho_z^I, \cr}
\qquad\qquad
\eqalign{
[Q,F_\zb^{\bar I}] & = -2\partial_\zb\chi^{\bar I}, \cr
[M,F_\zb^{\bar I}] & = 2\partial_\zb\chi^{\bar I}, \cr
[G_z,F_\zb^{\bar I}] & =\partial_z\rho_\zb^{\bar I} , \cr
[G_\zb,F_\zb^{\bar I}] & = 0. \cr}
\eqn\scinco
$$

Let us write the action corresponding to type A topological
matter in terms of the fields defined in \suno,
$$
\eqalign{
S=&{\nor}\int d^2z \Big[G_{I\bar J}\big(-F^I_z F^{\bar
J}_\zb - 2\rho_z^I D_\zb\chi^{\bar J} - 2
 \rho_\zb^{\bar J} D_z\chi^{I}   +
4\partial_z x^I \partial_\zb x^{\bar J}\big) \cr
&+\partial_K\partial_{\bar L} G_{I\bar J}
\rho_z^K \rho_\zb^{\bar L}\chi^I\chi^{\bar J}
-\partial_K G_{I\bar J} \chi^I F_\zb^{\bar J}\rho_z^K
-\partial_{\bar K} G_{I\bar J} \chi^{\bar K} F_z^{I}\rho_\zb^{\bar
J}\Big],\cr}
\eqn\cenueve
$$

where $G_{I\bar J}$ is the metric of the Kahler manifold $M$,
$$
G_{I\bar J}= { \partial^2 K \over \partial x^I \partial x^{\bar J}},
\eqn\cediez
$$
and $D_\mu$ represents a covariant derivative on sections
of the pull-back of the tangent bundle,
$$
D_\mu\chi^{I}  = \partial_\mu \chi^{I} +
                 (\partial_\mu x^J )\Gamma_{JK}^I \chi^{K},
\eqn\ceonce
$$
being $\Gamma_{JK}^I$  the Christoffel connection defined in (A22).
 Of course, the fields
$\rho_\mu^I$, $\rho_\mu^{\bar I}$, $F_\mu^I$ and
$F_\mu^{\bar I}$ entering \cenueve\ satisfy the selfduality and
anti-selfduality conditions,
$$
\eqalign{
\rho_\mu^I & = -i\epsilon_\mu{}^\nu \rho_\nu^I,
\,\,\,\,\,\,\,\,\,\,\,\,
\rho_\mu^{\bar I}  = i\epsilon_\mu{}^\nu \rho_\nu^{\bar I}, \cr
F_\mu^I & = -i\epsilon_\mu{}^\nu F_\nu^I,
\,\,\,\,\,\,\,\,\,\,\,\,
F_\mu^{\bar I}  = i\epsilon_\mu{}^\nu F_\nu^{\bar I}, \cr}
\eqn\ssiete
$$
where $\epsilon_\mu{}^\nu$ is such that $\epsilon_z{}^{z} =
-\epsilon_{\bar z}{}^{\bar z }= i$.

The action \cenueve\ possesses non-covariant looking terms which
can be reorganized into covariant looking ones performing a
redefinition of the auxiliary fields $F_\mu^I$ and $F_\mu^{\bar I}$
in such a way that all dependence on them becomes gaussian. This
is indeed the first step to carry out when integrating out those
fields. Let us define:
$$
\eqalign{
\tilde F_z^I =& F_z^I +\Gamma_{KJ}^I\chi^J\rho_z^K,\cr
\tilde F_\zb^{\bar I} =& F_\zb^{\bar I} +
\Gamma_{\bar K\bar J}^{\bar I}\chi^{\bar J}\rho_\zb^{\bar K}.\cr}
\eqn\extradi
$$
The action \cenueve\ becomes,
$$
S={\nor}\int d^2z \Big[G_{I\bar
J}\big( 4\partial_z x^I\partial_\zb x^{\bar J} - 2\rho_z^I
D_\zb\chi^{\bar J} - 2
 \rho_\zb^{\bar J} D_z\chi^{I}   - \tilde F_z^I \tilde F_\zb^{\bar J}
\big)
+ R_{\bar I J \bar K L} \rho_\zb^{\bar I}
\rho_z^J\chi^{\bar K}\chi^L \Big].
\eqn\extrade
$$

Notice the presence of a quartic term involving the curvature of
the Kahler manifold.
The redefinition  of auxiliary fields modifies the
symmetry transformations since they must be written in terms of the
new fields. Let us write down, for example, the form of the
$Q$-transformations,
$$
\eqalign{
[Q,x^I] & = \chi^I,\cr
\{Q,\chi^I\} & = 0,\cr
\{Q,\rho^I_z\} & =  \tilde F_z^I + 2\partial_z x^I - \Gamma_{JK}^I
\chi^J\rho_z^K,\cr
[Q,\tilde F_z^I] & =  -2D_z \chi^I
-\Gamma_{JK}^I\chi^J \tilde F_z^K
- R^I{}_{K\bar J L}\chi^K\chi^{\bar J}\rho_z^L,\cr
[Q,x^{\bar I}] & = \chi^{\bar I},\cr
\{Q,\chi^{\bar I}\} & = 0,\cr
\{Q,\rho^{\bar I}_\zb\} & =  \tilde F_\zb^{\bar I} + 2\partial_\zb x^{\bar
I} - \Gamma_{\bar J\bar K}^{\bar I} \chi^{\bar J}\rho_\zb^{\bar K},\cr
[Q,\tilde F_\zb^{\bar I}] & =  -2D_\zb \chi^{\bar I}
-\Gamma_{\bar J\bar K}^{\bar I}\chi^{\bar J} \tilde F_\zb^{\bar K}
+R^{\bar I}{}_{\bar J L\bar K}\chi^L\chi^{\bar J}\rho_\zb^{\bar K}. \cr
}  \eqn\extrado $$
Certainly, by construction, these transformations are such that
$Q^2=0$. Notice that according to \extrae\ and the definition
\extradi\ the $R$-transformations of the new auxiliary fields take
the same form as the old ones.

So far we have considered the theory on a flat two-dimensional
space. To analyze its topological character we must now place the
theory on an arbitrary curved two-dimensional manifold $\Sigma$ and verify
that it is still $Q$-invariant and that its energy-momentum tensor is
$Q$-exact. On an arbitrary two-dimensional manifold endowed with a metric
$g_{\mu\nu}$ the action \extrade\ takes the form:
$$
\eqalign{
S={\nor}\int_\Sigma d^2z\sqrt{g}\Big[&G_{I\bar
J}\big(  g^{\mu\nu}\partial_\mu x^I\partial_\nu x^{\bar J} +
{i\epsilon^{\mu\nu}\over \sqrt{g}} \partial_\mu x^I\partial_\nu x^{\bar J}
- g^{\mu\nu}\rho_\mu^I D_\nu\chi^{\bar J} \cr & -
 g^{\mu\nu}\rho_\mu^{\bar J} D_\nu\chi^{I}   - {1\over 2}g^{\mu\nu}
\tilde F_\mu^I \tilde
F_\nu^{\bar J} \big)
+{1\over 2}g^{\mu\nu} R_{\bar I J \bar K L} \rho_\mu^{\bar I}
\rho_\nu^J\chi^{\bar K}\chi^L \Big].\cr}
\eqn\extradu
$$

One can indeed verify that this action is invariant under the
covariantized form of the $Q$-transformations \extrado,
$$
\eqalign{
[Q,x^I] & = \chi^I,\cr
\{Q,\chi^I\} & = 0,\cr
\{Q,\rho^I_\mu\} & = \tilde F_\mu^I +  (\delta_\mu^\nu -
{i\epsilon_\mu{}^\nu \over \sqrt{g}})\partial_\nu x^I - \Gamma_{JK}^I
\chi^J\rho_\mu^K,\cr
[Q,\tilde F_\mu^I] & = - (\delta_\mu^\nu -
{i\epsilon_\mu{}^\nu \over \sqrt{g}})D_\nu \chi^I
-\Gamma_{JK}^I\chi^J \tilde F_\mu^K
- R^I{}_{K\bar J L}\chi^K\chi^{\bar J}\rho_\mu^L ,\cr
[Q,x^{\bar I}] & = \chi^{\bar I},\cr
\{Q,\chi^{\bar I}\} & = 0,\cr
\{Q,\rho^{\bar I}_\mu\} & = \tilde F_\mu^{\bar I} +
(\delta_\mu^\nu + {i\epsilon_\mu{}^\nu \over
\sqrt{g}})\partial_\nu x^{\bar I} - \Gamma_{\bar J\bar K}^{\bar I}
\chi^{\bar J}\rho_\mu^{\bar K},\cr
[Q,\tilde F_\mu^{\bar I}] & =
- (\delta_\mu^\nu + {i\epsilon_\mu{}^\nu \over \sqrt{g}})D_\nu
\chi^{\bar I} -\Gamma_{\bar J\bar K}^{\bar I}\chi^{\bar J} \tilde
F_\mu^{\bar K} + R^{\bar I}{}_{\bar J L\bar K}\chi^L\chi^{\bar
J}\rho_\mu^{\bar K}. \cr }
\eqn\extrafa
$$
Notice that the selfduality and anti-selfduality conditions \ssiete\ now
take the form,
$$
\eqalign{
\rho_\mu^I &= -i{\epsilon_\mu{}^\nu\over \sqrt{g}}\rho_\nu^I,\cr
F_\mu^I &= - i{\epsilon_\mu{}^\nu\over \sqrt{g}}F_\nu^I,\cr}
\qquad\qquad
\eqalign{
\rho_\mu^{\bar I} &= i{\epsilon_\mu{}^\nu \over\sqrt{g}}\rho_\nu^{\bar
I}, \cr
F_\mu^{\bar I}& = i{\epsilon_\mu{}^\nu \over\sqrt{g}}F_\nu^{\bar
I}. \cr}
\eqn\snueve
$$
The crucial test for the model that we have constructed being
topological is the verification of the $Q$-invariance of the action.
 It turns out that \extradu\ is indeed
invariant under the symmetry transformations generated by $Q$ and $M$ but it is
not invariant under the one generated by $G_\mu$. The invariance under $Q$ plus
the fact that the energy-momentum tensor can be written as a
$Q$-transformation,  guarantees that the model just constructed is a TQFT.
 Certainly, besides reparametrizations, the other
symmetries of the theory when considered on curved space are the $R$ and
$M$ symmetries. The  $R$-transformations are the ones given in \extrae. The
transformations under $M$ follow directly from the relation in \cocho,
$M=-2[R,Q]$. Notice that $Q\pm M$ generate independent nilpotent
symmetries for holomorphic and antiholomorphic modes. Recall that
according to \cseis\ they correspond to $Q_\ele$ and $Q_\ere$, \ie,
$Q_\ele = {1\over 2}(Q+M)$ and $Q_\ere= {1\over 2}(Q-M)$.

The construction that we have carried out guarantees that the
action of the theory is $Q$-exact. In order to have a TQFT we must check if a
stronger condition holds, namely, we must verify that the energy-momentum
tensor is $Q$-exact. As we will prove now we have a much stronger result
for these models of type A topological matter. It turns out that the action
itself is $Q$-exact. It is simple to demonstrate that,
$$
S=\Big\{Q, -{1\over 2}\int_\Sigma d^2z \sqrt{g}g^{\mu\nu}G_{I\bar
J}\big[{1\over 2} \rho_\mu^I {\tilde F}_\nu^{\bar J} +{1\over 2}
 \rho_\mu^{\bar J} {\tilde F}_\nu^I
-  ( \rho_\mu^I \partial_\nu x^{\bar J} + \rho_\mu^{\bar
J}\partial_\nu x^I) \big]\Big\},
\eqn\cedseis
$$
and therefore  the theory is certainly topological since this
implies that the energy-momentum is also $Q$-exact.

Once we have succeeded in the formulation of this topological quantum field
theory by twisting $N=2$ supersymmetry we may ask if it can be generalized.
All along our discussion the target space manifold ${M}$ was required
to be Kahler. Indeed, it is well known that $N=2$ supersymmetry requires a
Kahler manifold. However, after the twisting, one may ask if this condition
can be relaxed. Certainly, a field theory realization of the
part of the topological algebra \csiete\ which does not involve $G_\mu$ does
not
impose very restrictive conditions. The existence of a nilpotent operator
$Q$ is much weaker than the realization of the supersymmetry algebra in
which the anticommutator of two supersymmetric charges must correspond to a
vector operator. Thus in what regards to the $Q$ symmetry one could expect
a more general framework. The realization of the rest of the algebra
\csiete\ is only important when coupling this type of matter to
topological gravity. We will discuss that issue in sect. 5.

Topological sigma models as formulated by Witten [\wtsm] do indeed exist
for almost Hermitian manifolds. Let us  build its formulation from the one
we have obtained for Kahler manifolds. Since we are going to relax the
Kahler condition on ${M}$ we should first rewrite the theory changing
the holomorphic (antiholomorphic) notation in target space indices by
introducing a complex structure $J^i{}_j$, $i,j=1,...,2d,$.
For the case of a Kahler manifold this complex structure can be written
locally as: $J^I{}_J=-i\delta^I_J$, $I,J=1,...,d$; $J^{\bar I}{}_{\bar J}=
i\delta^{\bar I}_{\bar J}$,
${\bar I},{\bar J}=d+1,...,2d$; $J^{\bar I}{}_{J}=J^{J}{}_{\bar I}=0$,
 ${\bar I}=d+1,...,2d$,
$J=1,...,d$.
Notice that $J^i{}_kJ^k{}_j = -\delta_j^i$.
Thus, the action \extradu\ takes the form,
$$
\eqalign{
S={\nor}\int_\Sigma d^2z\sqrt{g}\Big[&\half G_{ij}g^{\mu\nu}
\partial_\mu x^i\partial_\nu x^{j}
+\half \varepsilon^{\mu\nu}J_{ij} \partial_\mu
x^i\partial_\nu x^{j}-G_{ij}g^{\mu\nu}\rho_\mu^i  D_\nu\chi^{j} \cr & -
{1\over 4}
 G_{ij}g^{\mu\nu}\tilde F_\mu^i\tilde F_\nu^{j} + {1\over 8} g^{\mu\nu}
R_{ijkm} \rho_\mu^{i} \rho_\nu^j\chi^{k}\chi^m \Big],\cr}
\eqn\luno
$$

and the selfduality and anti-selfduality conditions \snueve\ become,
$$
\eqalign{
\rho_\mu^i = \varepsilon_\mu{}^\nu J^i{}_j\rho_\nu^j,\cr}
\qquad\qquad
\eqalign{
\tilde F_\mu^i = \varepsilon_\mu{}^\nu J^i{}_j\tilde
F_\nu^j.\cr} \eqn\ldos
$$
In \luno\ and \ldos\ we have introduced the tensor,
$$
\varepsilon_{\mu\nu} = {\epsilon_{\mu\nu}\over \sqrt{g}}.
\eqn\ldosex
$$

So far we have only rewritten the action of type A topological matter in
compact notation with the help of a covariantly constant structure. Let us
know release this condition. Assume that the only requirements satisfied by
$J^i{}_j$ are
$$
J^i{}_kJ^k{}_j = -\delta_j^i,
\qquad\qquad
J^i{}_k J^j{}_m G_{ij} = G_{km},
\eqn\ltres
$$
\ie, $J^i{}_j$ is an almost-Hermitian structure. Notice that in a Kahler
manifold in addition to \ltres\ the condition $D_k J^i{}_j=0$ is satisfied.
Witten showed that the $Q$ transformations of the theory can be generalized
from the ones in \extrafa\ in such a way that the action \luno\ is
$Q$-invariant. Furthermore, this generalization is such that the nilpotency
of $Q$ holds. Actually, it is rather simple to obtain the form of these
$Q$-transformation from the ones given in \extrafa. First, let us rewrite
\extrafa\ in a compact form,
$$
\eqalign{
[Q,x^i] & = \chi^i,\cr
\{Q,\chi^i\} & = 0,\cr
\{Q,\rho^i_\mu\} & = \tilde F_\mu^i +  (\delta_\mu^\nu\delta^i_j +
\varepsilon_\mu{}^\nu J^i{}_j)\partial_\nu x^j -
\Gamma_{jk}^i \chi^j\rho_\mu^k,\cr
[Q,\tilde F_\mu^i] & =  -(\delta_\mu^\nu \delta^i_j +
\varepsilon_\mu{}^\nu J^i{}_j)D_\nu \chi^j
-\Gamma_{jk}^i\chi^j \tilde F_\mu^k
+{1\over 2} R^i{}_{kjm}\chi^k\chi^{j}\rho_\mu^m.\cr }
\eqn\lcuatro
$$
Certainly, the action \luno\ is invariant under
these transformations when the manifold is Kahler. In addition, for such a
case, the transformations are nilpotent. However, if the condition
$D_kJ^i{}_j=0$ is released none of the facts holds. One must modify the
transformations to achieve invariance and nilpotency. The procedure to
carry this out is simple. First, we will redefine the transformation in such
a way that $Q$ is nilpotent. Then, since the action was $Q$-exact for the
Kahler case, one may just take the compact version of \cedseis\ with the
new form of $Q$.
To redefine the transformations \lcuatro, notice that the first two does
not have to be modified. The $Q$-transformation of $\rho^i_\mu$, however,
has to be modified in such a way that the selfduality condition \dos\ is
maintained under the $Q$-transformation. It turns out that this is simple
to achieve by just adding a term of the form
$\varepsilon_{\mu}{}^\nu\chi^k\rho_\sigma^jD_kJ^i{}_j$. Once this is
obtained one fixes the transformation of $\tilde F_\mu^i$ in such a way
that $Q$ on $\rho^i_\mu$ is nilpotent. This leads to the following new set
of transformations:
$$
\eqalign{
[Q,x^i] =& \chi^i, \cr
\{Q,\chi^i\} =&0, \cr
\{Q,\rho_\mu^i\} =& \tilde F_\mu^i+\partial_\mu
x^i+\varepsilon_\mu{}^\nu J^i{}_j\partial_\nu x^j -
\Gamma_{jk}^i\chi^j\rho_\mu^k+ {1\over 2} \varepsilon_\mu{}^\nu \chi^k
\rho_\nu^j D_kJ^i{}_j,\cr
[Q,\tilde F^i_\mu] =& -D_\mu\chi^i -\varepsilon_\mu{}^\nu
 J^i{}_j
D_\nu\chi^j - \Gamma_{jk}^i\chi^j\tilde F_\mu^k +{1\over 2}
R_{mj}{}^i{}_k\chi^m\chi^j\rho_\mu^k \cr &
-\half\varepsilon_\mu{}^\nu
\chi^m\chi^k\rho_\nu^j D_mD_kJ^i{}_j
-{1\over 2}\varepsilon_\mu{}^\nu
 (D_k J^i{}_j)\chi^k
(\partial_\nu x^j -\varepsilon_\nu{}^\gamma
J^j{}_m\partial_\gamma x^m)\cr
& - {1\over 4}\chi^k\chi^m\rho_\mu^n (D_kJ^i{}_j)(D_mJ^j{}_n)
+{1\over 2} \varepsilon_\mu{}^\nu \chi^k\tilde F_\nu^j D_kJ^i{}_j. \cr}
\eqn\lcinco
$$
The crucial test for the validity of the construction resides on the fact
that $Q$ is also nilpotent on $\tilde F_\mu^i$. It is not obvious from
\lcinco\ that this is going to hold but an explicit computation shows
that, remarkably, it is true. This fact was discovered by Witten in
[\wtsm].

We have shown that the form of type A topological matter obtained after
twisting $N=2$ chiral multiplets can be generalized to the case of almost
Hermitian manifolds. The analysis of observables of this theory was carried
out in [\wtsm]. We will not discuss it here since our aim was to find out
which kind of supersymmetric matter leads to Witten's topological sigma
models. This analysis will provide the adequate framework to carry out the
coupling of this type of matter to topological gravity. This will be
discussed in sect. 5.

So far we have not been able to introduce potential terms for type A
topological matter. As we discussed at the beginning of this section,
F-terms are not allowed for the chiral multiplet since they lead to
actions which are not Lorentz invariant. There exists, however, some
$N=2$ supersymmetric models which can not be written in the form
\cicuatro\ and contain potential terms [\luis]. The twisting of these
models has been recently carried out [\nos] and it turns out that they
provide potential terms for type A topological matter for the case in
which the target manifold ${M}$ possesses some isometries.
In particular, the potential terms involve the Killing vectors associated
to those isometries. Let us
briefly review the results obtained in [\nos]. The
analysis presented in this paper clarifies the origin of eq. (2) in
[\nos], which might have seemed somehow obscure in that context.
The action presented there is just the twisted version of the general
$N=2$ supersymmetric action obtained in [\luis] with F-terms set to zero.
It turns out that regarding the $N=2$ supersymmetric multiplet as a chiral
multiplet, F-terms lead to non-Lorentz invariant expressions while terms
involving Killing vectors are permitted. The opposite occurs if one regards
the $N=2$ supersymmetric multiplet as twisted chiral. In components
fields, as it is the case of the construction given in [\luis],  one may have
one or the other multiplet depending on the $R$-charges which are assigned to
each of the members of the multiplet. In this construction, nothing guarantees
that the action that one obtains after the twisting can be generalized to the
almost Hermitian case since the construction in [\luis] assumes a Kahler
manifold with some isometries. As shown in [\nos] it turns out that,
indeed, it can be generalized to that case proceeding in a similar way as
in the case just described without potential terms.
Actually, in the process one finds a small surprise. The $N=2$ model under
consideration in [\luis] possesses central charges. This implies that $Q$
is not nilpotent any more. As shown in the previous section, there is no
need for $Q$ to be nilpotent to have a topological quantum field theory.
The only requirement is a $Q$ invariant action and an energy-momentum
tensor which is $Q$-exact. The model presented in [\nos] is such that
$Q^2$ is not zero but just a Lie derivative respect to some Killing vector
fields. We will summarize here the results presented in [\nos].

Let us consider an almost Hermitian manifold which possess two Killing
vector fields $V_i$ and $U_i$ which satisfy
$$
\eqalign{
D_i V_j + D_j V_i &=0,\cr
D_i U_j + D_j U_i &=0,\cr
V^j\partial_j U^i-U^j\partial_j V^i &=0.}
\qquad\qquad
\eqalign{
V^k\partial_kJ^i{}_n+J^i{}_k\partial_nV^k-J^k{}_n\partial_kV^i &=0,\cr
U^k\partial_kJ^i{}_n+J^i{}_k\partial_nU^k-J^k{}_n\partial_kU^i &=0,\cr}
\eqn\lseis
$$
These conditions are rather natural. They represent the requirement that
the metric and complex structure remain invariant under a variation along
the Killing vector fields. The most general action for type A topological
matter takes the form:
$$
\eqalign{
S=&\Big\{Q, \int_\Sigma \sqrt{g}\Big[
{1\over 2}g^{\mu\nu}G_{ij} \rho_\mu^i(\partial_\nu x^j
-{1\over 2}\tilde F^j_\nu)+ \lambda^2 G_{ij}(V^i+U^i)\chi^j\Big]\Big\}\cr
 =& \int_\Sigma \sqrt{g}
\Big[{1\over 2} G_{ij}g^{\mu\nu}\partial_\mu x^i \partial_\nu x^j
+{1\over 2} \varepsilon^{\mu\nu}J_{ij} \partial_\mu x^i
\partial_\nu x^j \cr
&- g^{\mu\nu}G_{ij}\rho^i_\mu \big(D_\nu\chi^j
+\half (D_k J_i{}^j)
\chi^k\varepsilon_\nu{}^\sigma
\partial_\sigma x^i\big)\cr
&-{1\over 4}g^{\mu\nu}\big(G_{ij}\tilde F^i_\mu \tilde F^j_\nu
-\half R_{ijkm}\rho^i_\mu\rho^j_\nu\chi^k\chi^m
+{1\over 4}(D_k J_{ip})(D_m J^p{}_j)
\rho^i_\mu\rho^j_\nu\chi^k\chi^m \big)\cr
&+\lambda^2 G_{ij}(V^iV^j - U^iU^j)
+ \lambda^2 \chi^i\chi^j D_i (V_j+U_j)
-{1\over 4} g^{\mu\nu}
\rho_\mu^i\rho^j_\nu D_i (V_j-U_j) \Big].\cr}
\eqn\lsiete
$$
The $Q$-transformations which leave this action invariant are the following,
$$
\eqalign{
[Q,x^i] =& \chi^i, \cr
\{Q,\chi^i\} =& (V^i-U^i), \cr
\{Q,\rho_\mu^i\} =& \tilde F_\mu^i+\partial_\mu x^i+\varepsilon_\mu{}^\nu
J^i{}_j\partial_\nu x^j - \Gamma_{jk}^i\chi^j\rho_\mu^k+
{1\over 2} \varepsilon_\mu{}^\nu \chi^k
\rho_\nu^j D_kJ^i{}_j,\cr
[Q,\tilde F^i_\mu] =& -D_\mu\chi^i -\varepsilon_\mu{}^\nu
 J^i{}_j
D_\nu\chi^j - \Gamma_{jk}^i\chi^j\tilde F_\mu^k +
{1\over 2} R_{mj}{}^i{}_k\chi^m\chi^j\rho_\mu^k \cr
&-\half\varepsilon_\mu{}^\nu
 \chi^m\chi^k\rho_\nu^j D_mD_kJ^i{}_j
-{1\over 2}\varepsilon_\mu{}^\nu (D_k J^i{}_j)\chi^k
(\partial_\nu x^j -\varepsilon_\nu{}^\sigma
J^j{}_m\partial_\sigma x^m)\cr
& - {1\over 4}\chi^k\chi^m\rho_\mu^n (D_kJ^i{}_j)(D_mJ^j{}_n)
+{1\over 2} \varepsilon_\mu{}^\nu \chi^k\tilde F_\nu^j D_kJ^i{}_j \cr
&+D_k (V^i-U^i) \rho_\mu^k - \half\varepsilon_\mu{}^\nu
(V^k-U^k)\rho_\nu^j
D_k J^i{}_j. \cr}
\eqn\locho
$$
Form these transformation one can easily verify that, indeed, $Q^2$ does
not vanish. It has the following action on the fields of the theory:
$$
\eqalign{
[Q^2, x^i] =& V^i-U^i, \cr
[Q^2, \chi^i] =& \partial_j (V^i-U^i) \chi^j, \cr
[Q^2, \rho_\mu^i] =& \partial_j (V^i-U^i) \rho_\mu^j, \cr
[Q^2, \tilde F_\mu^i] =& \partial_j (V^i-U^i) \tilde F_\mu^j, \cr}
\eqn\lnueve
$$
which just amounts to a Lie derivative respect to the Killing vector field
$V^i-U^i$.

The analysis of the observables of this theory was carried out in
[\nos]. They turn out to be the same as in the standard topological sigma
models [\wtsm] with the additional condition that the forms on  $M$ involved
are orthogonal to the difference of the Killing vector fields
$V^i-U^i$. Explicit computations of some observables were presented in [\nos].
For example, it was shown there that the partition function for the case in
which the two-dimensional manifold $\Sigma$ corresponds to the torus is just
the
Euler number of the target manifold $M$. This topological invariant was
obtained as the number of singular points of the Killing vector field.

\endpage

\chapter{\caps Type B Topological Matter}

The construction carried out in the previous section can be followed to
build theories involving topological matter of type B. We will briefly
describe here the steps involved in this construction. Our starting point
is a collection of twisted chiral superfields $X^I$ and  twisted
antichiral superfields $X^{\bar I}$, ($I,\bar I =1,...,d$),
$$
\eqalign{
D_{+,-} X^I & = 0, \cr
D_{+,+} X^{\bar I} & =0,\cr}
\qquad\qquad
\eqalign{
D_{-,+} X^I & = 0, \cr
D_{-,-} X^{\bar I} & =0. \cr}
\eqn\buno
$$
As in the previous case, our starting superspace action is \cicuatro.
However,  due to conditions \buno\ , the odd part of the
measure corresponding to the F-term now takes the following form,
$$
\eqalign{
d^2\theta &\rightarrow  D_{+,+}D_{-,-},\cr
\widehat{d^2\theta} &\rightarrow  D_{-,+}D_{+,-}.\cr }
\eqn\bdos
$$
These measures are certainly invariant under the new Lorentz generator
$\tilde J$. This means that
when twisting to obtain the corresponding TQFT we will have the
superpotential term. This is a remarkable difference respect to the case of
type A topological matter, where such terms were not allowed. Note on the
other hand that, as  announced, $R$-symmetry is preserved.

The next step in the construction is to define component fields and redefine
them as in \suno\ to make manifest their Lorentz structure respect to the new
Lorentz generator,  $\tilde J$, resulting after the twist. We define,
$$
\eqalign{
X^I| &=x^I, \cr
D_{+,+} X^I| & =\rho_z^I, \cr
D_{-,-} X^I |& = \rho_{\bar z}^I, \cr
D_{+,+} D_{-,-} X^I| &= F^I, \cr}
\qquad\qquad
\eqalign{
X^{\bar I}| &=x^{\bar I}, \cr
D_{+,-} X^{\bar I}| & = \chi^{\bar I}, \cr
D_{-,+} X^{\bar I}| & = \bar\chi^{\bar I}, \cr
D_{-,+} D_{+,-} X^{\bar I}| &= F^{\bar I}, \cr}
\eqn\btres
$$
Notice that, contrary to the case of type A topological matter, the field
$\rho^I_{\bar z}$ is not selfdual while the auxiliary fields $F^I$ and
$F^{\bar I}$ are scalars. The $R$-transformations of the fields appearing on
the right hand side of \btres\ are not manifiest any more. Let us collect them
here for later convenience,
$$
\eqalign{
[R,x^I] & = 0, \cr
[R,\rho^I_z] & = {1\over 2} \rho^I_z, \cr
[R,\rho^I_\zb] & = -{1\over 2}\rho^I_\zb,\cr
[R,F^I] & = 0.}
\qquad\qquad \eqalign{
[R,x^{\bar I}] & = 0, \cr
[R,\chi^{\bar I}] & = - {1\over 2} \chi^{\bar I}, \cr
[R,\bar{\chi}^{\bar I}] & = {1\over 2} \bar{\chi}^{\bar I}, \cr
[R,F^{\bar I}] & = 0. \cr}
\eqn\bseis
$$

The transformations of
the fields under the generators $Q$, $M$ and $G_\mu$ of the
topological algebra are easily obtained from \btres\ and the form of the
$N=2$ supersymmetric transformations \cinueve. They turn out to be,
$$
\eqalign{
[Q,x^I]& = 0, \cr
[M,x^I] & = 0, \cr
[G_z,x^I] & = {1\over 2}\rho_z^I, \cr
[G_\zb,x^I] & = {1\over 2}\rho_\zb^I,\cr}
\qquad\qquad
\eqalign{
[Q,x^{\bar I}] & = \chi^{\bar I} + \bar{\chi}^{\bar I}, \cr
[M,x^{\bar I}] & = \chi^{\bar I} - \bar{\chi}^{\bar I}, \cr
[G_z,x^{\bar I}] & = 0, \cr
[G_\zb,x^{\bar I}] & = 0,\cr}
\eqn\bsiete
$$
$$
\eqalign{
\{Q,\rho_z^I\} & = 2\partial_z x^I,\cr
\{M,\rho_z^I\} & = 2\partial_z x^I,\cr
\{G_z,\rho_z^I\} & = 0,\cr
\{G_\zb,\rho_z^I\} & = -{1\over 2}F^I,\cr}
\qquad\qquad
\eqalign{
\{Q,\rho_\zb^I\} & = 2\partial_\zb x^I,\cr
\{M,\rho_\zb^I\} & = -2\partial_\zb x^I,\cr
\{G_z,\rho_\zb^I\} & = {1\over 2}F^I,\cr
\{G_\zb,\rho_\zb^I\} & =0,\cr}
\eqn\bocho
$$
$$
\eqalign{
\{Q,\chi^{\bar I}\} & = F^{\bar I},\cr
\{M,\chi^{\bar I}\} & = -F^{\bar I},\cr
\{G_z,\chi^{\bar I}\} & = \partial_z x^{\bar I},\cr
\{G_\zb,\chi^{\bar I}\} & = 0,\cr}
\qquad\qquad
\eqalign{
\{Q,\bar{\chi}^{\bar I}\} & = -F^{\bar I},\cr
\{M,\bar{\chi}^{\bar I}\} & = -F^{\bar I},\cr
\{G_z,\bar{\chi}^{\bar I}\} & = 0,\cr
\{G_\zb,\bar{\chi}^{\bar I}\} & = \partial_\zb x^{\bar I},\cr}
\eqn\bnueve
$$
$$
\eqalign{
[Q,F^I] & = 2\partial_z\rho_\zb^I - 2\partial_\zb\rho_z^I, \cr
[M,F^I] & = 2\partial_z\rho_\zb^I + 2\partial_\zb\rho_z^I, \cr
[G_z,F^I] & = 0, \cr
[G_\zb,F^I] & = 0, \cr}
\qquad\qquad
\eqalign{
[Q,F^{\bar I}] & = 0, \cr
[M,F^{\bar I}] & = 0, \cr
[G_z,F^{\bar I}] & = -\partial_z\bar{\chi}^{\bar I}, \cr
[G_\zb,F^{\bar I}] & = \partial_\zb\chi^{\bar I}. \cr}
\eqn\bdiez
$$

The action resulting from \cicuatro\ after using the definitions \btres\ takes
the following form,
$$
\eqalign{
S'={\norp}\int d^2z & \Big[G_{I\bar J}\big(F^I F^{\bar
J} +2 \rho_z^I D_\zb\chi^{\bar J} +
 2\rho_\zb^{I} D_z\bar {\chi}^{\bar J}   -
4\partial_\zb x^I \partial_z x^{\bar J}\big) \cr
&+\partial_K\partial_{\bar L} G_{I\bar J}
\bar {\chi}^{\bar L}\rho_z^K \rho_\zb^{I}\chi^{\bar J}
+\partial_K G_{I\bar J}\rho_z^K \rho_\zb^{I}  F^{\bar J}
+\partial_{\bar K} G_{I\bar J} \bar {\chi}^{\bar K} F^{I}\chi^{\bar J} \cr
&+(\partial_J\partial_I W)\rho_z^J \rho_\zb^{I}+ (\partial_I W) F^{I}
-(\partial_{\bar J}\partial_{\bar I} \bar W)\chi^{\bar J}\bar {\chi}^{\bar I}
+ (\partial_{\bar I} \bar W) F^{\bar I}
\Big].\cr}
\eqn\bonce
$$

Notice the presence of potential terms in this action.
For the moment we have denoted this action by $S'$ since it will be
convenient to redefine it by a global factor later. We will reserve the symbol
$S$ for its final form.  The auxiliary fields $F^I$ and $F^{\bar I}$ can be
integrated out in the usual way. Furthermore, as in the previous case, this
procedure makes the action manifiestly reparametrization invariant from the
point of view of Kahler geometry. We will reduce the dependence of the action
on  $F^I$ and $F^{\bar I}$ to a simple quadratic term plus other terms
involving the potentials $W$ and $\bar W$ carrying out the following
definition,
$$
\eqalign{
\tilde F^I =& F^I +\Gamma_{JK}^I\rho_z^J\rho_\zb^K,\cr
\tilde F^{\bar I} =& F^{\bar I} +
\Gamma_{\bar J\bar K}^{\bar I}\bar {\chi}^{\bar J}\chi^{\bar K},\cr}
\eqn\bdoce
$$
where $\Gamma_{JK}^I$ and $\Gamma_{\bar J\bar K}^{\bar I}$ are the Christoffel
connections defined in (A22). The action \bonce\ becomes,
$$
\eqalign{
S'={\norp}\int d^2z & \Big[G_{I\bar
J}\big( -4\partial_\zb x^I\partial_z x^{\bar J}+
2  \rho_z^I D_\zb\chi^{\bar J} +
 2  \rho_\zb^{I} D_z\bar {\chi}^{\bar J}
+ \tilde F^I \tilde F^{\bar J}
\big) \cr
& + R_{ I\bar L  K\bar J} \rho_\zb^{I}
\bar {\chi}^{\bar L}\rho_z^{K}\chi^{\bar J}
+(D_I\partial_J W)\rho_z^I \rho_\zb^{J}+ (\partial_I W) \tilde
F^{I} \cr
&-(D_{\bar I}\partial_{\bar J}\bar W)\chi^{\bar I}\bar {\chi}^{\bar J} +
(\partial_{\bar I} \bar W) \tilde F^{\bar I}
 \Big].\cr}
\eqn\btrece
$$

So far we have carried out the twisting procedure. Now we are in the position
of verifying if the theory is topological. To this end we must
first place it on an arbitrary two dimensional manifold, and verify that the
action  is $Q$-invariant and  the energy-momentum tensor is $Q$-exact. Let us
therefore consider a Riemann surface $\Sigma$ endowed with a metric
$g_{\mu\nu}$. The covariantization of the action \btrece\ takes the form,
$$
\eqalign{
S'=\int_\Sigma d^2\sigma &\sqrt{g}\Big[  G_{I\bar J}
\big(- g^{\mu\nu}
  \partial_\mu x^I\partial_\nu
  x^{\bar J} + i\varepsilon^{\mu\nu}\partial_\mu x^I\partial_\nu
  x^{\bar J} + {1\over 2}g^{\mu\nu}
\rho_{\mu}^I D_\nu(\chi^{\bar
  J} +\bar\chi^{\bar J}) \cr &
+ {i\over 2}\varepsilon^{\mu\nu}\rho_{\mu}^{I}
D_{\nu}(\chi^{\bar J}
  -{\bar\chi}^{\bar J})
+ \tilde F^I \tilde F^{\bar J}\big)
-{i\over 4}\varepsilon^{\mu\nu}
R_{I\bar L  K\bar J} \rho_{\mu}^{I} \bar {\chi}^{\bar
  L}\rho_{\nu}^{K}\chi^{\bar J} \cr &
+ (\partial_I W) \tilde F^{I}
  -(D_{\bar I}\partial_{\bar J}\bar W)\chi^{\bar I}\bar {\chi}^{\bar J} +
  (\partial_{\bar I} \bar W) \tilde F^{\bar I}
+{i\over 4}\varepsilon^{\mu\nu}
  (D_I\partial_J W)\rho_{\mu}^I\rho_{\nu}^{J} \Big],\cr}
\eqn\bcatorce
$$
where we have used \ldosex.
Notice that this action has a minus sign in the term
$- g^{\mu\nu} \partial_\mu x^I\partial_\nu x^{\bar J}$ as compared to
\extradu. This is reminiscent of a well known fact in $N=2$ supersymmetry,
 where chiral and twisted chiral multiplets provide kinetic terms with
opposite signs [\roc]. If one identifies $G_{I\bar J}$ with a positive
definite metric one should have to change the global sign of the whole
action to have a bosonic part leading to a convergent functional
integral. Notice that although the bosonic part of the action is not
real, its imaginary part is a topological invariant. From now on we will
assume that $G_{I\bar J}$ is positive definite and we will introduce a global
negative sign to the action \bcatorce. The resulting action turns out to be,
$$
\eqalign{
S=\int_\Sigma d^2\sigma &\sqrt{g}\Big[  G_{I\bar J}
\big( g^{\mu\nu}
  \partial_\mu x^I\partial_\nu
  x^{\bar J} - i\varepsilon^{\mu\nu}\partial_\mu x^I\partial_\nu
  x^{\bar J} - {1\over 2}g^{\mu\nu}
\rho_{\mu}^I D_\nu(\chi^{\bar
  J} +\bar\chi^{\bar J}) \cr &
- {i\over 2} \varepsilon^{\mu\nu}\rho_{\mu}^{I}
D_{\nu}(\chi^{\bar J}
  -{\bar\chi}^{\bar J})
- \tilde F^I \tilde F^{\bar J}\big)
+{i\over 4}\varepsilon^{\mu\nu}
R_{I\bar L  K\bar J} \rho_{\mu}^{I} \bar {\chi}^{\bar
  L}\rho_{\nu}^{K}\chi^{\bar J} \cr &
- (\partial_I W) \tilde F^{I}
  +(D_{\bar I}\partial_{\bar J}\bar W)\chi^{\bar I}\bar {\chi}^{\bar J} -
  (\partial_{\bar I} \bar W) \tilde F^{\bar I}
-{i\over 4}\varepsilon^{\mu\nu}
  (D_I\partial_J W)\rho_{\mu}^I\rho_{\nu}^{J} \Big].\cr}
\eqn\bcatorcep
$$

Taking into account the
redefinitions \bdoce, one easily derives the covariantized form of the
$Q$-transformations of the fields,
$$
 \eqalign{ [Q,x^I]& = 0, \cr [Q,x^{\bar
I}] & = \chi^{\bar I} + \bar{\chi}^{\bar I}, \cr \{Q,\rho_{\mu}^I\} & =
2\partial_{\mu} x^I,\cr \{Q,\chi^{\bar I}\} & = \tilde F^{\bar I}
  -\Gamma_{\bar J\bar K}^{\bar I}\bar {\chi}^{\bar J}\chi^{\bar K},\cr
\{Q,\bar {\chi}^{\bar I}\} & = -\tilde F^{\bar I}
  +\Gamma_{\bar J\bar K}^{\bar I}\bar {\chi}^{\bar J}\chi^{\bar K},\cr
[Q,\tilde F^I] & = i\varepsilon^{\mu\nu} \big[
D_{\mu}\rho_{\nu}^I
  +{1\over 4} R_{J \bar T K}^I(\chi^{\bar T}+
  \bar {\chi}^{\bar T})\rho_{\mu}^J\rho_{\nu}^K \big] , \cr
[Q,\tilde F^{\bar I}] & = \Gamma_{\bar J\bar K}^{\bar I}
  \tilde F^{\bar J}(\bar {\chi}^{\bar K} +\chi^{\bar K}).\cr}
\eqn\bquince
$$
It is simple to verify that, indeed, the action is invariant under this set of
transformations. Furthermore, similarly to the case of type A topological
matter, one can verify that the action is also invariant under $M$ and $R$
transformations. However, it is not invariant under transformations
generated by $G_\mu$. Before carrying out the analysis of the energy-momentum
tensor one could ask if the action \bcatorcep\ is $Q$-exact as in the
previous case. The answer to this question is negative. Only when the
potential terms are not present the action is  $Q$-exact. Let us set
$W=0$ in \bcatorcep. It is simple to verify that the remaining action
can be written as,
$$
S|_{W=0}=\Big\{Q,\int_\Sigma d^2\sigma \sqrt{g}\Big[  G_{I\bar J}
\big({1\over
2} g^{\mu\nu}\rho_\mu^I\partial_\nu x^{\bar J} - {i\over 2}
\varepsilon^{\mu\nu} \rho_\mu^I\partial_\nu x^{\bar J}-
\tilde F^I\chi^{\bar J}\big)
\Big]\Big\}.
\eqn\bdseis
$$
When potential terms are present, the action
\bcatorcep\ is not $Q$-exact. A short analysis shows that the transformations
\bquince\ can not generate a term like $(\partial_IW)\tilde F^I$ as the one
present in the action \bcatorcep. However, the requirement for the theory being
topological is just that the energy-momentum tensor be $Q$-exact. The
energy-momentum can be easily computed from the variation of the action
\bcatorcep\ respect to the two-dimensional metric. One finds:
$$
\eqalign{
T_{\mu\nu} = &{1\over 2} (\delta_{\mu}^{\sigma}\delta_\nu^\tau
+\delta_{\mu}^{\tau}\delta_\nu^\sigma)G_{I\bar J}
\big(\partial_\sigma x^I\partial_\tau x^{\bar J} - {1\over 2}
\rho_\sigma^I D_\tau (\chi^{\bar J} + \bar \chi^{\bar J})\big) \cr &
-{1\over 2} g_{\mu\nu}\Big[G_{I\bar J}\Big(g^{\sigma\tau}
\big(\partial_\sigma x^I\partial_\tau x^{\bar J} - {1\over 2}
\rho_\sigma^I D_\tau (\chi^{\bar J} + \bar \chi^{\bar J})\big)
-\tilde F^I\tilde F^{\bar J}\Big) \cr &
- \partial_IW \tilde F^I
-\partial_{\bar I} \bar W \tilde F^{\bar I} + (D_{\bar I}
\partial_{\bar J} \bar W)\chi^{\bar I}\bar \chi^{\bar J}\Big]. \cr}
\eqn\bdsiete
$$
Notice that, again, one has the same problem as before since the term
$(\partial_IW)\tilde F^I$ is present. Since the field $\tilde F^I$ is
the source of the problem and, on the other hand, it is auxiliary, we will
integrate it out. The price to pay in doing this is that the algebra will not
close off-shell, \ie, we will have $Q^2=0$ modulo  field equations.
However, if we succeed in proving that after integrating out the auxiliary
field the energy-momentum tensor of the theory is $Q$-exact we will have shown
that the theory is topological. The integration of the auxiliary field will
give some dependence on the two dimensional metric but this can be factorized
and the rest must correspond to a topological invariant.
In other words, here is no need of an off-shell realization
for the theory being topological since the possible metric dependence
originated from the auxiliary fields can be factorized.
Let us define,
$$
\hat F^I = \tilde F^I + G^{I\bar J}\partial_{\bar J}\bar W
\qquad\qquad
\hat F^{\bar I} = \tilde F^{\bar I} + G^{\bar I J}\partial_{J} W.
\eqn\bdsietep
$$

The dependence of the action \bcatorcep\ on the auxiliary fields $\hat F^I$ and
$\hat F^{\bar I}$ becomes gaussian. The integration of the auxiliary fields
$\hat F^I$ and $\hat F^{\bar I}$ gives a dependence on the two dimensional
metric which can be factorized from the functional integral. For example, the
integration of the constant auxiliary modes gives a factor which is
proportional to the volume of the two dimensional manifold to some power. The
resulting action turns out to be,
 $$
\eqalign{
S=\int_\Sigma d^2\sigma &\sqrt{g}\Big[  G_{I\bar J}
\big( g^{\mu\nu}
  \partial_\mu x^I\partial_\nu
  x^{\bar J} - i\varepsilon^{\mu\nu}\partial_\mu x^I\partial_\nu
  x^{\bar J} - {1\over 2}g^{\mu\nu}
\rho_{\mu}^I D_\nu(\chi^{\bar
  J} +\bar\chi^{\bar J}) \cr &
- {i\over 2}\varepsilon^{\mu\nu}\rho_{\mu}^{I}
D_{\nu}(\chi^{\bar J}
  -{\bar\chi}^{\bar J}) \big)
+{i\over 4}\varepsilon^{\mu\nu}
R_{I\bar L  K\bar J} \rho_{\mu}^{I} \bar {\chi}^{\bar
  L}\rho_{\nu}^{K}\chi^{\bar J} \cr &
+ G^{I\bar J}(\partial_I W) \partial_{\bar J}\bar W
  +(D_{\bar I}\partial_{\bar J}\bar W)\chi^{\bar I}\bar {\chi}^{\bar J}
-{i\over 4}\varepsilon^{\mu\nu}
  (D_I\partial_J W)\rho_{\mu}^I\rho_{\nu}^{J} \Big].\cr}
\eqn\bdocho
$$
Now we have to check if the
 theory possess an energy-momentum tensor which is $Q$-exact.
After using \bdsietep\ and setting $F^I=F^{\bar I}=0$ (since they have been
integrated out) the transformations \bquince\ become,
$$ \eqalign{ [Q,x^I]& = 0, \cr
[Q,x^{\bar I}] & = \chi^{\bar I} + \bar{\chi}^{\bar I}, \cr
\{Q,\rho_{\mu}^I\} & = 2\partial_{\mu} x^I,\cr
\{Q,\chi^{\bar I}\} & = -G^{\bar I J}\partial_J W
  -\Gamma_{\bar J\bar K}^{\bar I}\bar {\chi}^{\bar J}\chi^{\bar K},\cr
\{Q,\bar {\chi}^{\bar I}\} & = G^{\bar I J}\partial_J W
  +\Gamma_{\bar J\bar K}^{\bar I}\bar {\chi}^{\bar J}\chi^{\bar K}.\cr}
\eqn\bdnueve
$$
Using these transformations one finds that the energy-momentum tensor
corresponding to the action \bdocho\ is $Q$-exact,
$$
\eqalign{
T_{\mu\nu} =  \Big\{ Q, &{1\over 2} g_{\mu\nu}\big(g^{\sigma\tau}G_{I\bar
J}\rho_\sigma^I\partial_\tau x^{\bar J} - (\chi^{\bar I}-\bar\chi^{\bar
I})\partial_{\bar I} \bar W\big) \cr
&\,\,\,\,\,\,\,\, -{1\over 2}G_{I\bar J}(\rho_\mu^I\partial_\nu x^{\bar J}
+\rho_\nu^I\partial_\mu x^{\bar J})\Big\} \cr}
\eqn\bveinte
$$
For the
case in which the target space is flat, the theory we have constructed was
first
studied in [\vafa].

The appearance of a topological quantum field theory where the $Q$-symmetry is
only realized on-shell is not new. A classical example where this also occurs
is topological Yang-Mills theory in four dimensions \REF\witDP{E.
Witten\journal\cmp&117(88)353} [\witDP]. However, the theory we have
constructed possess a feature which is not present in topological Yang-Mills
theory in four dimensions. Namely, the action \bdocho\ is not $Q$-exact. In
this
sense the theory we have constructed has also features of other classes of
topological quantum field theories as Chern-Simons theory in three dimensions
\REF\wics{E. Witten\journal\cmp&121(89)351} [\wics].

Let us construct the observables corresponding to type B topological matter.
First we will build the ones corresponding to zero forms and then, we will use
\veintep\ to obtain their descendants. This construction will show the
usefulness of the operator $G_\mu$ in dealing with observables.
{}From the $Q$-transformations of the fields \bdnueve\ follows
trivially that any function which depends only on $x^I$ and not on $x^{\bar
I}$, \ie, holomorphic from the point of view of the target space, is
$Q$-invariant and therefore an observable. Then, in the notation used in
\veintep\ and \vseis, we have $$
\phi^{(0)} = A(x^I).
\eqn\bvuno
$$
Using the operator $G_\mu$, whose action on the fields is given in
\bsiete, \bocho, \bnueve\ and \bdiez, one finds,
$$
\eqalign{
\phi^{(1)}_\mu =& [G_\mu, \phi^{(0)}] = {1\over 2}\partial_I
A\rho_\mu^I, \cr
\phi^{(2)}_{\mu\nu} = & {1\over 2} \{ G_\mu, \phi^{(1)}_\nu\} =
{1\over 8}D_J\partial_I A \rho_\mu^J\rho_\nu^I -
{i\over 4}\varepsilon_{\mu\nu} \partial_I A
(\hat F^I - \partial^I\bar W).\cr}
\eqn\bvdos
$$
Notice that we have restored all the  dependence on the auxiliary fields in
computing \bvdos. The reason for this is that only off-shell the topological
algebra holds. Otherwise it holds modulo field equations and then the analysis
of observables is more complicated. The topological algebra \csiete\
guarantees that the $Q$-transformations of the fields in \bvdos\ leads to a
total derivative and therefore their integration over closed 1-cycles and
2-cycles respectively leads to observables. One can check this explicitly
using the transformations listed in \bsiete, \bocho, \bnueve\ and \bdiez.
Furthermore, as shown in \vnueve, to obtain a possible non-vanishing vacuum
expectation value one must consider closed 1-cycles which are
homologically non-trivial. The only 2-cycle to be consider is the
two-dimensional manifolds itself. The two-form operator \bvdos\ possesses a
feature which is not standard in topological quantum field theories, namely, it
depends on the auxiliary field $\hat F^I$. Since, as discussed above, the
presence of the auxiliary field leads to an energy-momentum tensor which is not
$Q$-exact, we have to reanalyze the type of invariants obtained from operators
as  $\phi^{(2)}_{\mu\nu}$ in \bvdos. Before carrying out such analysis we will
construct the second type of $Q$-invariant operators of the theory.

Let us consider a closed form of type $(0,p)$, \ie, a closed form with $p$
antiholomorphic indices, $A_{\bar I_1 \bar I_2 ... \bar I_p}$. Certainly, the
operators
$$
\tilde\phi^{(0)} = A_{\bar I_1 \bar I_2 ... \bar I_p}
(\chi^{\bar I_1}+\bar \chi^{\bar I_1})
(\chi^{\bar I_2}+\bar \chi^{\bar I_2})...
(\chi^{\bar I_p}+\bar \chi^{\bar I_p})
\eqn\bvtres
$$
are $Q$-invariant. The construction of their descendants is carried out using
the operator $G_\mu$ as before. One finds,
$$
\eqalign{
\tilde\phi^{(1)}_\mu =&{1\over 2}\partial_J
 A_{\bar I_1 \bar I_2 ... \bar I_p}\rho^J_\mu
(\chi^{\bar I_1}+\bar \chi^{\bar I_1})
(\chi^{\bar I_2}+\bar \chi^{\bar I_2})...
(\chi^{\bar I_p}+\bar \chi^{\bar I_p}) \cr
&+  A_{\bar I_1 \bar I_2 ... \bar I_p}
\sum_{s=1}^p (-1)^{s+1} (\chi^{\bar I_1}+\bar \chi^{\bar I_1})
...\partial_\mu x^{\bar I_s}
...(\chi^{\bar I_p}+\bar
\chi^{\bar I_p}), \cr
\tilde\phi^{(2)}_{\mu\nu} =&{1\over 8}D_J\partial_K
A_{\bar I_1 \bar I_2 ... \bar I_p}\rho^J_\mu\rho^K_\nu
(\chi^{\bar I_1}+\bar \chi^{\bar I_1})
(\chi^{\bar I_2}+\bar \chi^{\bar I_2})...
(\chi^{\bar I_p}+\bar \chi^{\bar I_p}) \cr
&-{i\over 4}\varepsilon_{\mu\nu}
\partial_J A_{\bar I_1 \bar I_2 ... \bar I_p}(\hat F^J-
\partial^{J}\bar W)(\chi^{\bar I_1}+\bar \chi^{\bar I_1})
(\chi^{\bar I_2}+\bar \chi^{\bar I_2})...
(\chi^{\bar I_p}+\bar \chi^{\bar I_p}) \cr
&+{1\over 4}\partial_J
A_{\bar I_1 \bar I_2 ... \bar I_p}\rho_{[\mu}^J
\sum_{s=1}^p(-1)^{s+1}
(\chi^{\bar I_1}+\bar \chi^{\bar I_1})...
\partial_{\nu]} x^{\bar I_s}...
(\chi^{\bar I_p}+\bar \chi^{\bar I_p}) \cr
&+{1\over 2}A_{\bar I_1 \bar I_2 ... \bar I_p}
\sum_{s,t=1\atop s\neq t}^p(-1)^{t+s}
(\chi^{\bar I_1}+\bar \chi^{\bar I_1})...
\partial_{\mu} x^{\bar I_s}...
\partial_{\nu} x^{\bar I_t}...
(\chi^{\bar I_p}+\bar \chi^{\bar I_p}), \cr}
\eqn\bvcuatro
$$
where the squared brackets on indices denote antisymmetrization with no factor.
Again, as in the previous operators, we observe the same feature regarding the
dependence on the auxiliary field. Let us analyze the consecuences of having
a linear dependence on $F^I$ in \bvdos\ and \bvcuatro.

We will denote by $\varphi_{\mu\nu}^{(2)}$ ($\tilde\varphi_{\mu\nu}^{(2)}$)
the part of $\phi_{\mu\nu}^{(2)}$ in \bvdos\ ($\tilde\phi_{\mu\nu}^{(2)}$ in
\bvcuatro) which does not contain $\hat F^I$. Ones has,
$$
\eqalign{
\phi_{\mu\nu}^{(2)} =& \varphi_{\mu\nu}^{(2)} +
{i\over 4} \varepsilon_{\mu\nu}F^I\xi_I, \cr \tilde\phi_{\mu\nu}^{(2)}
=& \tilde\varphi_{\mu\nu}^{(2)} + {i\over 4}\varepsilon_{\mu\nu}
F^I\tilde\xi_I,\cr} \eqn\bvcinco
$$
where,
$$
\eqalign{
\xi_I = & - \partial_IA,\cr
\tilde\xi_I = &
-\partial_I A_{\bar I_1 \bar I_2 ... \bar I_p}(\chi^{\bar I_1}+\bar
\chi^{\bar I_1}) (\chi^{\bar I_2}+\bar \chi^{\bar I_2})...
(\chi^{\bar I_p}+\bar \chi^{\bar I_p}). \cr}
\eqn\bvseis
$$
We will show that the vacuum expectation values of the integrated form
of the operators
 $\varphi_{\mu\nu}^{(2)}$ and  $\tilde\varphi_{\mu\nu}^{(2)}$ computed with
the action \bdocho\ are topological invariants. Namely, we will prove,
$$
\eqalign{
{\delta\over\delta g^{\mu\nu}}
\langle\int_\Sigma \varphi_{\mu\nu}^{(2)} \rangle =& 0, \cr
{\delta\over\delta g^{\mu\nu}}
\langle\int_\Sigma \tilde\varphi_{\mu\nu}^{(2)} \rangle =& 0. \cr}
\eqn\bvsiete
$$
To prove this we will place back the auxiliary fields in the action of the
theory and we will compute the vacuum expectation value of the integrated form
of the operators $\phi_{\mu\nu}^{(2)}$ and  $\tilde\phi_{\mu\nu}^{(2)}$. In
other words we will consider the theory off-shell. We will carry out
 the analysis explicitly for $\phi_{\mu\nu}^{(2)}$ but, similarly, it follows
for $\tilde\phi_{\mu\nu}^{(2)}$ since the only fact that we need to use is
that the dependence on $F^I$ is linear and this is a common feature to both
operators. Let us therefore consider
$$
\langle\int_\Sigma \phi_{\mu\nu}^{(2)} \rangle_{\hbox{\sevenrm off}} =
\int [dX][dF]\int_\Sigma(\varphi_{\mu\nu}^{(2)} +
{i\over 4}\varepsilon_{\mu\nu} \hat
F^I\xi_I)\ex^{-S(X)+\int_\Sigma\sqrt{g} G_{I\bar J} \hat F^I \hat F^{\bar J}},
\eqn\bvocho
$$
where $[dX]$ denotes the measure corresponding to the fields
$x^I$, $x^{\bar I}$, $\chi^{\bar I}$, $\bar\chi^{\bar I}$,
$\rho_\mu{}^I$, and $[dF]$ the one corresponding to $\hat F^I$ and
$\hat F^{\bar I}$. The subindex  in $\langle\rangle_{\hbox{\sevenrm off}}$
 denotes that the vacuum expectation value is taken off-shell.
Since the dependence of  $\phi_{\mu\nu}^{(2)}$ on $F^I$ is linear one has,
$$
\eqalign{
\langle\int_\Sigma \phi_{\mu\nu}^{(2)} \rangle_{\hbox{\sevenrm off}} = &
\int [dX][dF]\int_\Sigma\varphi_{\mu\nu}^{(2)}
\ex^{-S(X)+\int_\Sigma\sqrt{g}
G_{I\bar J} \hat F^I \hat F^{\bar J}}\cr
= & \int [dF] \ex^{\int_\Sigma\sqrt{g}
G_{I\bar J} \hat F^I \hat F^{\bar J}}
\langle \int_{\Sigma} \varphi_{\mu\nu}^{(2)} \rangle.\cr}
\eqn\bvnueve
$$
This result shows that the vacuum expectation value \bvocho\ factorizes in a
part which contains the integration on $F^I$ and it is not topological
invariant, times the vacuum expectation value $\langle \int_{\Sigma}
\varphi_{\mu\nu}^{(2)} \rangle$, where the functional integration is carried
out without the fields $F^I$ and $F^{\bar I}$. Our aim is to show that this
last factor is topological invariant. To carry this out we will take the vacuum
expectation value \bvocho\ and we will study its dependence on $g_{\mu\nu}$. We
have, $$
\eqalign{
{1\over\sqrt{g}}{\delta \over \delta g^{\sigma\tau}}
& \langle \int_{\Sigma} \phi_{\mu\nu}^{(2)} \rangle_{\hbox{\sevenrm off}}
 \cr
& = \int [dX][dF]\int_\Sigma\phi_{\mu\nu}^{(2)}
(T_{\sigma\tau} + {1\over 2}g_{\sigma\tau}G_{I\bar J}\hat F^I \hat F^{\bar J})
\ex^{-S(X)+\int_\Sigma\sqrt{g}
G_{I\bar J} \hat F^I \hat F^{\bar J}}\cr
&=  \int [dX][dF] \{ Q,\int_\Sigma\phi_{\mu\nu}^{(2)}G_{\sigma\tau} \}
\ex^{-S(X)+\int_\Sigma\sqrt{g}
G_{I\bar J} \hat F^I \hat F^{\bar J}} \cr
&{\hbox{\hskip0.5cm}}+ {1\over 2}g_{\sigma\tau} \int [dX][dF]
\phi_{\mu\nu}^{(2)}  G_{I\bar J} \hat F^I \hat F^{\bar J}
\ex^{-S(X)+\int_\Sigma\sqrt{g}
G_{I\bar J} \hat F^I \hat F^{\bar J}}, \cr}
\eqn\btreinta
$$
where we have used  that $[Q,\phi_{\mu\nu}^{(2)}]$ is a total derivative
and we have denoted by $G_{\sigma\tau}$ the quantity such that $T_{\sigma\tau}=
\{Q,G_{\sigma\tau}\}$ in \bveinte\ (recall eq. \tres). The first term in
\btreinta\ vanishes because, certainly, the off-shell action which appears in
the exponent is $Q$-invariant. We are left with the second term which can be
written as, $$
{1\over\sqrt{g}}{\delta \over \delta g^{\sigma\tau}}
\langle \int_{\Sigma} \phi_{\mu\nu}^{(2)} \rangle_{\hbox{\sevenrm off}}
=  - {1\over 2}g_{\sigma\tau} \int [dF] G_{I\bar J} \hat F^I \hat F^{\bar J}
\ex^{\int_\Sigma\sqrt{g}
G_{I\bar J} \hat F^I \hat F^{\bar J}}
\langle \int_{\Sigma} \varphi_{\mu\nu}^{(2)} \rangle,
\eqn\btuno
$$
since the linear dependence on $F^I$ in $\phi_{\mu\nu}^{(2)}$ gives a
vanishing contribution. Comparing \btuno\ with \bvnueve\ one concludes that,
indeed,
$$
{1\over\sqrt{g}}{\delta \over \delta g^{\sigma\tau}}
\langle \int_{\Sigma} \varphi_{\mu\nu}^{(2)} \rangle = 0.
\eqn\btdos
$$
This result proves that the on-shell theory leads to topological invariants.
Notice that it has been essential in the proof that the dependence on $F^I$ of
the observables is at most linear in the auxiliary fields.

\endpage

\chapter{\caps Coupling to Topological Gravity}

In this section we will describe a gauging procedure to build the coupling
 of the theories involving topological matter constructed in the two previous
sections  to topological gravity. The two types of topological quantum field
theories which we have constructed possess the $Q$-symmetries generated by the
transformations \locho\ and \bdnueve. If one considers a flat two-dimensional
space it is clear from the construction that the actions \lsiete\ and \bdocho\
of the two types of theories are invariant under $G_\mu$-transformations. When
considering the theories on a curved two-dimensional space, however, these
actions are not $G_\mu$-invariant. The approach that we will describe in this
section to couple topological matter to topological gravity will consists  of a
modification of these theories in such a way that the resulting theories are
invariant under a {\it local} $G_\mu$-symmetry. This will be done by
introducing a  gauge field corresponding to this symmetry which we will denote
by $\psi_{\mu\nu}$. Certainly, since $G_\mu$ is odd, this new gauge field is
also odd. From the relation $\{Q,G_\mu\}=P_\mu$ in \csiete\ follows that the
gauge field associated to $G_\mu$ must be the $Q$-partner of the metric
$g_{\mu\nu}$, which is the gauge field associated to $P_\mu$. We will find out
in the construction that, indeed, invariance of the gauged action under $Q$
implies that $g_{\mu\nu}$ and $\psi_{\mu\nu}$ are $Q$-partners, \ie,
$[Q,g_{\mu\nu}]=\psi_{\mu\nu}$. Notice that in this construction $P_\mu$ and
$G_{\mu}$ are generators of local symmetries while $Q$ is a global one which
relates both.

In this paper we will consider the coupling of a simple type B model. A
treatment in full generality will be presented elsewhere. Let us consider type
B topological matter corresponding to a flat $2d$-dimensional target space
with no potential terms. The corresponding action is easily obtained from
\bdocho,
 $$
\eqalign{
S=\int_\Sigma d^2\sigma \sqrt{g}
\big( & g^{\mu\nu}
  \partial_\mu x^I\partial_\nu
  x^{\bar I} - i\varepsilon^{\mu\nu}\partial_\mu x^I\partial_\nu
  x^{\bar I} \cr & - {1\over 2}g^{\mu\nu}
\rho_{\mu}^I \partial_\nu(\chi^{\bar
  I} +\bar\chi^{\bar I})
- {i\over 2}\varepsilon^{\mu\nu}\rho_{\mu}^{I}
\partial_{\nu}(\chi^{\bar I}
  -{\bar\chi}^{\bar I}) \big). \cr}
\eqn\bttres
$$
This action is invariant under the $Q$-transformations \bdnueve\ that now take
the form,
$$ \eqalign{ [Q,x^I]& = 0, \cr
[Q,x^{\bar I}] & = \chi^{\bar I} + \bar{\chi}^{\bar I}, \cr
\{Q,\rho_{\mu}^I\} & = 2\partial_{\mu} x^I.\cr}
\qquad\qquad
\eqalign{
\{Q,\chi^{\bar I}\} & = 0,\cr
\{Q,\bar {\chi}^{\bar I}\} & = 0,\cr}
\eqn\btcuatro
$$
However, \bttres\ is not invariant under the $G_\mu$-transformation which can
be easily obtained from  \bsiete, \bocho\ and \bnueve,
$$
\eqalign{
[G_\mu,x^I] = & {1\over 2} \rho_\mu^I,\cr
[G_\mu,x^{\bar I}] = & 0,\cr
\{G_\mu,\chi^{\bar I}\} = &{1\over 2}(\partial_\mu x^{\bar I}
-i\varepsilon_\mu{}^\nu\partial_\nu x^{\bar I}),\cr
\{G_\mu,\bar\chi^{\bar I}\} = &{1\over 2}(\partial_\mu x^{\bar I}
+i\varepsilon_\mu{}^\nu\partial_\nu x^{\bar I}),\cr
\{G_\mu,\rho_\nu^I\} = & 0.\cr}
\eqn\btcinco
$$

The action \bttres\ is reparametrization invariant. On the other hand, the
$G_\mu$-transformations \btcinco\ are covariant. Since the action \bttres\ is
$G_\mu$-invariant for a flat two-dimensional space, it must be also invariant
if the parameter associated to a $G_\mu$-transformation is covariantly
constant. Let us introduce an odd local parameter $\eta^\mu$ as the one
corresponding to $G_\mu$-transformations. From \btcinco\ these transformations
can be written as,
$$
\eqalign{
\delta x^I = & {1\over 2} \eta^\mu\rho_\mu^I, \cr
\delta x^{\bar I} = & 0, \cr
\delta \chi^{\bar I} = &{1\over 2}\eta^\mu(\partial_\mu x^{\bar I}
-i\varepsilon_\mu{}^\nu\partial_\nu x^{\bar I}),\cr
\delta\bar\chi^{\bar I} = &{1\over 2}\eta^\mu(\partial_\mu x^{\bar I}
+i\varepsilon_\mu{}^\nu\partial_\nu x^{\bar I}),\cr
\delta\rho_\nu^I = & 0.\cr}
\eqn\btseis
$$
The variation of the action \bttres\ under these transformations takes the
form,
$$
\delta S = \int_\Sigma d^2\sigma \sqrt{g}
(\nabla^\mu\eta^\nu)P_{\mu\nu}{}^{\sigma\tau}\rho^I_\sigma\partial_\tau x^{\bar
I}, \eqn\btsiete
$$
where $P_{\mu\nu}{}^{\sigma\tau}$ is a projector into the traceless symmetric
part for tensors of rank two,
$$
P_{\mu\nu}{}^{\sigma\tau} =  {1\over 2}
(\delta_\sigma^\mu\delta_\tau^\nu + \delta_\sigma^\nu\delta_\tau^\mu
-g_{\mu\nu}g^{\sigma\tau}),
\eqn\btocho
$$
and $\nabla_\mu$ is a two-dimensional covariant derivative.
Notice that, as expected, the variation \btsiete\ vanishes for a covariantly
constant parameter $\eta^\mu$.

Our next step is the introduction of a new odd field $\psi_{\mu\nu}$ which
will play the role of gauge field for the transformations \btseis. From the
variation \btsiete\ follows that this field must be symmetric and traceless,
$$
\psi_{\mu\nu} = \psi_{\nu\mu},
\qquad\qquad
\psi_\mu{}^\mu = 0,
\eqn\btnueve
$$
and must transform as,
$$
\delta \psi_{\mu\nu} = 2P_{\mu\nu}{}^{\sigma\tau}\nabla_\sigma\eta_\tau.
\eqn\bcuarenta
$$
The term to be added to the action \bttres\ must be such that the action is
invariant under the transformations \btseis\ and \bcuarenta. This term is
simple to guess. The gauged action turns out to be,
 $$
\eqalign{
S_{\hbox{\sevenrm g}}=\int_\Sigma  & d^2\sigma \sqrt{g}
\big(  g^{\mu\nu}
  \partial_\mu x^I\partial_\nu
  x^{\bar I} - i\varepsilon^{\mu\nu}\partial_\mu x^I\partial_\nu
  x^{\bar I} \cr & - {1\over 2}g^{\mu\nu}
\rho_{\mu}^I \partial_\nu(\chi^{\bar
  I} +\bar\chi^{\bar I})
- {i\over 2}\varepsilon^{\mu\nu}\rho_{\mu}^{I}
\partial_{\nu}(\chi^{\bar I}
  -{\bar\chi}^{\bar I})
-{1\over 2}\psi^{\mu\nu}\rho_\mu^I\partial_\nu x^{\bar I} \big). \cr}
\eqn\bcuno
$$
Notice that the metric tensor $g_{\mu\nu}$ does not transform under $G_\mu$
transformations. The $Q$-variation of the action \bcuno\ is not defined since
we have not specified the $Q$-transformation of $\psi_{\mu\nu}$. Furthermore,
so far we have consider $g_{\mu\nu}$  as a $Q$-invariant quantity. It turns out
that the only way to make \bcuno\ $Q$-invariant is defining the
$Q$-transformations of $g_{\mu\nu}$ and $\psi_{\mu\nu}$ as,
$$
\eqalign{
[Q,g_{\mu\nu}] = & \psi_{\mu\nu}, \cr
\{Q,\psi_{\mu\nu}\} = & 0, \cr}
\eqn\bcdos
$$
which are consistent with the nilpotency of $Q$. The gauge action is invariant
under ordinary reparametrizations and the local symmetry listed in
\btseis\ and \bcuarenta. To quantize the gauged action \bcuno\ one needs to
fix these local gauge symmetries. These gauge fixings leads to the introduction
of ghost fields which build the standard content of topological gravity in two
dimensions  [\lpw,\ms,\ver]. We will not describe the quantization procedure
in this work. It follows the lines described in [\lpw,\ver].

In this section we have coupled topological matter to topological gravity
taking a simple model for type B matter. The procedure should be extended to
the general case. The steps needed in the gauging procedure are rather
standard and one does not expect unsurmountable complications. We expect to
report on this in  the future.

\endpage

\chapter{\caps Concluding remarks}

In this work we have constructed two types of theories containing topological
matter after twisting $N=2$ supersymmetry. In addition we have given a gauging
procedure to couple matter to topological gravity.
Type A topological matter is a topological quantum field theory whose action
is $Q$-invariant. However, this feature is not shared by type B
topological matter. It should be desirable to understand the nature of the
observables associated to  type B topological matter and, in particular, the
role played by the potential which, certainly, is going to be non-trivial. One
should also study the relation between these theories and conformal topological
field theories. For example, it should be interesting to analyze if a relation
as the one between supersymmetric Landau-Ginzburg models and N=2 superconformal
models \REF\mart{E. Martinec\journal\pl&B217(89)431; {\it Criticality,
Catastrophe and Compactifications,} V.G. Knizhnik memorial volume, 1989}
\REF\vawar{C. Vafa and N.P. Warner\journal\pl&B218(89)51}
\REF\lvw{W. Lerche, C. Vafa and N.P. Warner\journal\np&B324(89)427}
[\mart,\vawar,\lvw] holds. It is not
clear in such a picture which one is the correspondence between the observables
constructed in sect. 4 for type B matter and the ones in topological conformal
field theory [\ey,\li].

It is interesting to remark that associated to the two types of topological
matter that we have studied there also exist their conjugate counterparts. The
choice of new Lorentz generator \dos, $\tilde J = J+R$ to carry out the twist
could have been chosen differently, namely,
$\tilde J = J-R$. It is clear from the construction that the resulting
theories would have the same features as the ones constructed in sect. 3 and
4. In type A theories one would obtain a change of selfduality by
anti-selfduality conditions. In type B theories the prominent role played by
the anti-holomorphic coordinates will be played by the holomorphic ones.

Theories containing a mixture of type A and type B topological matter should be
constructed. It would very interesting to analyze its geometric interpretation
as well as the possible potential terms which they allow. The resulting models
must be related to the supersymmetric ones constructed in [\roc] and therefore
they will provide a topological version of Wess-Zumino-Witten models.
 Finally, the full coupling to topological gravity of all these types of
topological matter should be carried out.

\vskip1cm

\ack
We would like to thank L. Alvarez-Gaum\'e, J. Mas, A.V. Ramallo and J.
S\'anchez-Guill\'en  for helpful discussions. P. M. Llatas would
like to thank  the
Departamento de F\'\i sica de Part\'\i culas da Universidade de Santiago,
where this work was carried out, for its hospitality.
This work was supported in part by DGICYT under grant PB90-0772, and by CICYT
under grants AEN88-0013 and AEN88-0040.
\endpage

\appendix


In this appendix we will give a summary of our conventions and we will recall
a few facts about complex manifolds. The quantum field theories considered in
this paper are defined on a two-dimensional manifold  $\Sigma$ which is locally
endowed with an  Euclidean metric $g_{\mu\nu}=\delta_{\mu\nu}$. This metric
is invariant under $SO(2)$ tangent space rotations which generate the
corresponding ``Lorentz" group. Greek indices from the beginning of
the alphabet label spinor representations of the Lorentz group
while greek indices of the middle of the alphabet label vector
representations.   A real system of coordinates on
$\Sigma$ is denoted by $x^1$, $x^2$ which combine to give holomorphic
coordinates,
$$
z=x^1 + ix^2, \,\,\,\,\,  {\bar z}=x^1 - ix^2,
\eqn\aauno
$$
in such a way that a vector $V^\mu$ with real components $(V^1,V^2)$ has
holomorphic components given by,
$$
V_z={1\over 2}(V_1 - iV_2),
\,\,\,\,\,\,\,\,\,\,  V_{\bar z}={1\over 2}(V_1 +iV_2) .
\eqn\aados
$$
The components of the locally flat Euclidean metric are:
$$
\eqalign{
g_{z\bar z}=&g_{{\bar z}z}={1\over 2},\cr
g_{zz}=&g_{{\bar z}\bar z}=0, \cr}
\qquad\qquad
\eqalign{
g^{z\bar z}=&g^{{\bar z}z}=2, \cr
g^{zz}=&g^{{\bar z}\bar z}=0.\cr}
\eqn\aatres
$$
The epsilon symbol is is chosen in such a way that,
$$
\epsilon^{12}=-\epsilon^{21}=1,
\eqn\aacuatro
$$
or, in holomorphic coordinates,
$$
\epsilon^{{\bar z}z}=-\epsilon^{z{\bar z}}=2i.
\eqn\aacinco
$$
On a curved space, the epsilon symbol behaves as a tensor density in such a
way that $\varepsilon^{\mu\nu}$ defined by,
$$
\varepsilon^{\mu\nu}={1\over{\sqrt{g}}}\epsilon^{\mu\nu}
\eqn\aaseis
$$
behaves as a tensor.


Our choice of  Euclidean Dirac matrices $\gamma^\mu$ is,
$$
(\gamma^1)_\alpha{}^{\beta} =\sigma^1, \qquad\qquad
(\gamma^2)_\alpha{}^{{}\beta} =\sigma^2,
\eqn\aanueve
$$
where $\sigma^1$, $\sigma^2$ are Pauli matrices. Lorentz
spinor indices are lowered and raised by the  matrix
$C_{\alpha\beta}=\sigma^1$,
$$
(\gamma^\mu)_{\alpha\beta}=(\gamma^\mu)_\alpha{}^\tau C_{\tau\beta}.
\eqn\aadiez
$$

The algebra of the generators of $N=2$ supersymmetry takes the form,
$$
\{ Q_{\alpha}{}^a,Q_{{\beta}b}\} =\delta_b{}^a\gamma^{\mu}_{\alpha\beta}P_\mu.
\eqn\aaonce
$$
where Latin indices label the spinor representation of the internal $SO(2)$
symmetry.  When both, Lorentz
$SO(2)$ and internal $SO(2)$ indices are written explicitly, they are
separated by a comma. The algebra \aaonce\ reads:
$$
\eqalign{
\{ Q_{+,+}\,\, ,\,\, Q_{+,-} \} =& 2P_z =2\partial_z,\cr
\{ Q_{-,+}\,\, ,\,\, Q_{-,-} \} =& 2P_{\bar z} =2\partial_{\bar z},\cr}
\eqn\aadoce
$$
while all others anticommutators vanish.

Let us recall a few facts about complex geometry which are useful  for the
comprehension of the paper. Let us consider even-dimensional manifold $M$.
 An $almost$ $complex$ $structure$ $J_j{}^i$ on $M$ is a $(1,1)$ tensor
satisfying,
$$
J_i{}^k J_k{}^j=-\delta_i^j.  \eqn\auno
$$
A manifold $M$ is called $almost$ $complex$ if it admits an
almost complex structure $J_i{}^j$.
This manifold $M$ is called a $complex$ $manifold$ if it admits
an atlas with a complex coordinate system in such a way that transition
functions between neighbor charts are holomorphic. In other words, if the
complex coordinates of two patches $U$ and $V$ ($U\cap V\not=\emptyset$) are
$X^I$, $X^{\bar I},$ and $Y^I$, $Y^{\bar I},$ respectively, then the transition
functions in $U\cap V$ are such that,
$$
\eqalign{
X^I =& X^I (Y^I), \,\,\,\,\,\,\,\,\,\, \partial_{Y^{\bar I}} X^I =0,\cr
X^{\bar I} =& X^{\bar I} (Y^{\bar I}), \,\,\,\,\,\,\,\,\,\, \partial_{Y^I}
X^{\bar I} =0.\cr}
 \eqn\ados
$$
In this holomorphic complex coordinate system the complex structure
$J^i{}_j$ can be defined with constant entries,
$$
J_I{}^J =-i\delta_I^J, \,\,\,\,\,\,\ J_{\bar I}{}^{\bar J}=i\delta_{\bar
I}^{\bar J}, \,\,\,\,\,\,\ J_I{}^{\bar J} = J_{\bar I}{}^J=0. \eqn\atres
$$

A complex manifold with Riemannian metric $G_{ij}$ is called $almost$
$Hermitian$ if
$$
G_{ij}=J_i^p J_j^l G_{pl}. \eqn\acuatro
$$
Using property \auno\ this statement is equivalent to,
$$
J_{ij}\equiv J_i{}^k G_{kj} = -J_{ji}. \eqn\acinco
$$
Hermiticity is a restriction on the metric, and not on the manifold. If
a complex manifold admits a metric $H_{ij}$, then it also admits the
metric $G_{ij}$ defined as,
$$
G_{ij}={1\over 2}(H_{ij}+J_i{}^p J_j{}^l H_{pl}), \eqn\aseis
$$
which is Hermitian. Moreover, in  holomorphic coordinates, after using  \atres,
$$
G_{IJ}=G_{{\bar I}{\bar J}}=0, \eqn\asiete
$$
being the nonvanishing components of the metric  of the type
$G_{I{\bar J}}=G_{{\bar J}I}$.

{}From \acinco\ follows that  on an almost Hermitian manifold one can naturally
define the two-form $J$,
$$
J={1\over 2} J_{ij} dX^i \wedge dX^j. \eqn\aocho
$$
An almost Hermitian manifold is called $Kahler$ if this two-form is closed,
$$
dJ=0, \eqn\anueve
$$
which, in holomorphic components becomes,
$$
\partial_K G_{I{\bar J}}=\partial_I G_{K{\bar J}}, \,\,\,\,\,\,\,\
\partial_{\bar K} G_{I{\bar J}}=\partial_{\bar J} G_{I{\bar K}}. \eqn\adiez
$$
This implies the existence of a $Kahler$ $potential$ $K(X^I,X^{\bar I})$ such
that,
$$
G_{I{\bar J}}=\partial_I \partial_{\bar J} K.  \eqn\aonce
$$
Christoffel connections can be computed straightforwardly from previous
expressions. The nonvanishing components are:
$$
\Gamma_{JK}^I = G^{I\bar L}\partial_J G_{K \bar
L},\,\,\,\,\,\,\,\,\,\, \Gamma_{\bar J\bar K}^{\bar I} = G^{\bar I
L}\partial_{\bar J} G_{L \bar K}.
\eqn\adoce
$$
This fact implies that
the only non-trivial components of the Riemann tensor are,
$$
R_{\bar I J \bar K L} =
G_{\bar I M} \partial_{\bar K}\Gamma^M_{JL},
\eqn\atrece
$$
plus all others which are obtained using the
symmetry properties of the Riemann tensor.

\endpage

\refout
\end